\documentclass[a4paper]{scrartcl}
\pdfoutput=1
\usepackage[utf8]{inputenc}

\usepackage[T1]{fontenc} 
\usepackage{textcomp} 
\usepackage{lmodern} 
\usepackage[sc,osf]{mathpazo} 
\usepackage{myT1classico} 
\DeclareMathAlphabet{\mathsf}{T1}
  {\sfdefault}{m}{n} 
\SetMathAlphabet{\mathsf}{bold}{T1}{\sfdefault}{b}{n} 

\usepackage{amsmath,amssymb,amsthm}
\usepackage{mathtools} 
\usepackage{tensor}
\usepackage{accents}
\usepackage{cancel}

\usepackage{enumitem}
\usepackage{relsize}
\newcommand{\acronym}[1]{\texorpdfstring{\textsmaller{#1}}{#1}}

\usepackage[british]{babel}

\usepackage{microtype} 

\usepackage{authblk}

\usepackage[colorlinks]{hyperref}

\DeclareFontFamily{U}{matha}{\hyphenchar\font45}
\DeclareFontShape{U}{matha}{m}{n}{
  <5> <6> <7> <8> <9> <10> gen * matha
  <10.95> matha10 <12> <14.4> <17.28> <20.74> <24.88> matha12
}{}
\DeclareSymbolFont{matha}{U}{matha}{m}{n}
\DeclareMathSymbol{\oleft}{2}{matha}{"68}
\DeclareMathSymbol{\oright}{2}{matha}{"69}

\usepackage[numbers,sort&compress]{natbib}
\makeatletter
\@ifundefined{overunderset}{
  \newcommand{\overunderset}[3]{\binrel@{#3}%
    \binrel@@{\mathop{\kern\z@#3}\limits^{#1}_{#2}}}
}
\makeatother

\DeclareMathOperator{\Or}{O} 
\newcommand{\e}{\mathrm{e}} 
\newcommand{\E}{\mathrm{E}} 
\newcommand{\I}{\mathrm{i}} 
\newcommand{\D}{\mathrm{d}} 
\newcommand{\DD}{\mathrm{D}} 

\newcommand{\bomega}{{\boldsymbol{\omega}}}
\newcommand{\btheta}{{\boldsymbol{\theta}}}
\newcommand{\bhatomega}{{\boldsymbol{\hat\omega}}}
\newcommand{\bTheta}{{\boldsymbol{\Theta}}}
\newcommand{\ba}{{\boldsymbol{a}}}
\newcommand{\R}{\mathbb R}

\newcommand{\Aut}{\mathsf{Aut}}
\newcommand{\Diff}{\mathsf{Diff}}
\newcommand{\GL}{\mathsf{GL}}
\newcommand{\On}{\mathsf{O}}
\newcommand{\Gal}{\mathsf{Gal}}
\newcommand{\IGal}{\mathsf{IGal}}
\newcommand{\Barg}{\mathsf{Barg}}
\newcommand{\Ui}{\mathsf{U}(1)}

\newcommand{\Der}{\mathsf{Der}}
\newcommand{\gl}{\mathfrak{gl}}
\newcommand{\so}{\mathfrak{so}}
\newcommand{\gal}{\mathfrak{gal}}
\newcommand{\ui}{\mathfrak{u}(1)}

\theoremstyle{plain}
\newtheorem{theorem}{Theorem}[section]

\numberwithin{equation}{section}

\title{Teleparallel Newton--Cartan gravity}

\author{Philip K. Schwartz}
\affil{Institute for Theoretical Physics,
  Leibniz University Hannover, \par
  Appelstraße 2, 30167 Hannover, Germany}
\affil{\normalfont\texttt{\href{mailto:philip.schwartz@itp.uni-hannover.de}
    {philip.schwartz@itp.uni-hannover.de}}}

\date{}

\begin{document}
\maketitle

\begin{abstract}
  \noindent
  We discuss a teleparallel version of Newton--Cartan gravity.  This
  theory arises as a formal large-speed-of-light limit of the
  teleparallel equivalent of general relativity (\acronym{TEGR}).
  Thus, it provides a geometric formulation of the Newtonian limit of
  \acronym{TEGR}, similar to standard Newton--Cartan gravity being the
  Newtonian limit of general relativity.  We show how by a certain
  gauge-fixing the standard formulation of Newtonian gravity can be
  recovered.
\end{abstract}

\section{Introduction}

Any viable theory of gravity has to reproduce Newtonian gravity in the
slow-velocity, weak-gravity Newtonian limit.  For standard general
relativity (\acronym{GR}), this Newtonian limit is traditionally
exhibited by linearising the Einstein equations around a Minkowski
spacetime background and imposing small velocities of matter,
employing a suitable choice of coordinates (see, e.g., the textbook
accounts \cite[chapter 18]{MTW:2017}, \cite[section
5.1]{Straumann:2013}).

However, there exists also a geometric, coordinate-free formulation of
this limiting process: standard Newtonian gravity may be reformulated
in geometric terms as so-called \emph{Newton--Cartan gravity}, which
features a Galilei-relativistic spacetime geometry and describes
Newtonian gravity by a curved connection, similar to \acronym{GR}.
Newton--Cartan gravity can be shown to arise as the Newtonian limit of
\acronym{GR}.  It was originally developed by Cartan in 1923
\cite{Cartan:1923,Cartan:1924} and independently by Friedrichs in 1926
(published in 1928) \cite{Friedrichs:1928}.  Important further
contributions to Newton--Cartan gravity were made by Trautman in the
1960s \cite{Trautman:1963,Trautman:1965}, by Künzle in the 1970s
\cite{Kuenzle:1972,Kuenzle:1976}, and by Ehlers in a 1981 article
\cite{Ehlers:1981a,Ehlers:1981b} on the precise relation between
\acronym{GR} and Newton--Cartan gravity.  An elementary textbook
account of Newton--Cartan gravity is given in a book on foundational
issues of \acronym{GR} by Malament \cite[chapter 4]{Malament:2012}, to
which we also refer for further historical references.

In this article, we answer the question what the corresponding
geometric description of the Newtonian limit looks like for the
teleparallel equivalent of general relativity (\acronym{TEGR})
\cite{Maluf:2013}.  Teleparallel gravity \cite{Bahamonde.EtAl:2021} is
a class of gravitational theories in which the geometric framework for
the description of gravity is modified compared to that of standard
\acronym{GR}: in addition to the spacetime metric, one considers as
basic ingredient a connection that is metric but, differently to the
Levi-Civita connection, torsionful and flat.  This framework allows
for a dynamically equivalent formulation of \acronym{GR}, namely the
above-mentioned \acronym{TEGR}, and modified teleparallel theories of
gravity have recently gained a lot of interest in the context of
classical modified gravity or effective descriptions of quantum
gravity \cite{Bahamonde.EtAl:2021,CANTATA:2021,
  Bahamonde.Ducobu.Pfeifer:2022,Bahamonde.EtAl:2022}.

The geometry of Galilei manifolds, which underlies Newton--Cartan
gravity, admits a `gauge-theoretic' description in terms of the
Bargmann group (i.e.\ the centrally extended inhomogeneous Galilei
group), which was initially discovered in the 1980s in the context of
massive matter coupling \cite{Duval.Kuenzle:1984,Duval.EtAl:1985}.
Over the last decade, interest in Galilei geometries has significantly
increased due to novel applications in condensed matter physics as
well as in `non-relativistic' string theory and related applications
in quantum gravity \cite{Son:2013, Andringa.EtAl:2012,
  Christensen.EtAl:2014a, Christensen.EtAl:2014b,
  Bergshoeff.EtAl:2022}.  In this context, the gauge-theoretic
perspective has been further developed \cite{Andringa.EtAl:2011,
  Geracie.EtAl:2015}.  It is this perspective on the geometry of
Galilei manifolds in which a teleparallel formulation of
Newton--Cartan gravity arises very naturally, which turns out to give
the sought-for geometric description of the Newtonian limit of
\acronym{TEGR}.

For the consideration of modifications of \acronym{GR}, there are some
typical physical motivations, such as the problem of quantisation,
observational issues in cosmology or astrophysics, or the prediction
of singularities.  Of those motivations, none applies to Newtonian
gravity---for example, Newton--Cartan gravity coupled to a matter
field can be rigorously quantised \cite{Christian:1997}.  Therefore,
one might argue that on physical grounds, the investigation of
geometrically modified descriptions of Newtonian gravity just by
itself does not seem necessary.  Notwithstanding this argument, our
investigation of a `teleparallelised' version of Newton--Cartan
gravity that gives the geometric description of the Newtonian limit of
teleparallel gravity is still of fundamental importance on a both
conceptual and theoretical level: the description of a physically
interesting limit of an inherently geometric theory ought to be given
in a geometric fashion.

A teleparallel formulation of Newton--Cartan gravity was also
constructed by Read and Teh in \cite{Read.Teh:2018}.  Their approach
is different from ours in two important aspects.  First, in
\cite{Read.Teh:2018} the theory is constructed only in a restricted,
`gauge-fixed' situation; in the present paper, we develop instead a
completely general teleparallel description of Newton--Cartan gravity,
without introducing arbitrary assumptions on the connection or the
frame.  (In fact, the formalism of \cite{Read.Teh:2018} may be shown
to arise as a special case of ours.)  Second, \cite{Read.Teh:2018}
does not analyse the Newtonian limit of \acronym{TEGR}, but instead
shows that its teleparallel formulation of Newton--Cartan gravity can
be obtained from higher-dimensional \acronym{TEGR} by null reduction,
i.e.\ quotienting along a lightlike symmetry \cite{Duval.EtAl:1985,
  Julia.Nicolai:1995}.

The structure of this paper is as follows.  First, we quickly
introduce notation and conventions used throughout the article in
section \ref{subsec:notation}.  In section
\ref{sec:Barg_structs_telep_Gal_conns}, we introduce the geometric
framework for our formulation of teleparallel Newton--Cartan gravity
(starting with a review of the geometry of Galilei manifolds), and
formulate the theory.  Section \ref{sec:telepar_NC_from_TEGR} shows
that this theory is the Newtonian $c\to\infty$ limit of
\acronym{TEGR}; in section \ref{sec:recover_Newton} we discuss how to
recover the standard formulation of Newtonian gravity from
teleparallel Newton--Cartan gravity.  We conclude and discuss possible
directions for future research in section \ref{sec:conclusion}.

\subsection{Notation and conventions}
\label{subsec:notation}

\paragraph{Lie groups and algebras}

For a Lie group denoted by a Latin capital letter (or by several
letters), the corresponding Lie algebra will be denoted by the
corresponding lowercase Fraktur letter, e.g.\ $\mathfrak{g} =
\mathrm{Lie}(G)$.

The semidirect product of Lie groups $H$ and $N$ with respect to a
homomorphism $\rho \colon H \to \Aut(N)$ will be denoted by
\begin{equation}
  H \ltimes_\rho N.
\end{equation}
Similarly, the semidirect sum of Lie algebras $\mathfrak{h}$ and
$\mathfrak{n}$ with respect to a homomorphism $\tilde\rho \colon
\mathfrak{h} \to \Der(\mathfrak{n})$ will be denoted by
\begin{equation}
  \mathfrak{h} \oright_{\tilde\rho} \mathfrak{n}.
\end{equation}
As a vector space, this is of course just the direct sum of
$\mathfrak{h}$ and $\mathfrak{n}$, but we want to take account of the
nontrivial Lie bracket structure in the notation.  In the case of both
semidirect products and sums, if the homomorphism is clear from
context we will omit it from the notation.

\paragraph{The Galilei and Bargmann groups}

When we speak of the Galilei group without further qualification, we
will mean the orthochronous homogeneous Galilei group, which is a
semidirect product
\begin{equation}
  \Gal = \On(3) \ltimes \R^3
\end{equation}
whose parts are interpreted as spatial (improper) rotations and
Galilei boosts, respectively, where the homomorphism for the
semidirect product is the defining representation of
$\On(3)$.\footnote{Note that the restriction to the orthochronous
  Galilei group is not essential at all; including time reversal would
  just introduce the necessity to define the notion of Galilei
  manifolds in section \ref{sec:Barg_structs_telep_Gal_conns} in terms
  of a time metric instead of a clock one-form, which automatically
  introduces a choice of time orientation (and time-orientability).}

The inhomogeneous (orthochronous) Galilei group is the semidirect
product
\begin{equation}
  \IGal = \Gal \ltimes \R^4
\end{equation}
of homogeneous transformations and spacetime translations, where the
homomorphism $\Gal \to \Aut(\R^4)$ is given by
\begin{subequations} \label{eq:action_Gal_R4}
\begin{equation}
  (R,k)(s,y) = (s, Ry + sk)
\end{equation}
for $(R,k) \in \Gal$ and $(s,y) \in \R \times \R^3 = \R^4$.  We can
thus view $\Gal$ as a subgroup of $\GL(4)$, with the element $(R,k)
\in \Gal$ corresponding to the matrix
\begin{equation}
  \begin{pmatrix}
    1 & 0 \\
    k & R
  \end{pmatrix}.
\end{equation}
\end{subequations}

The Bargmann group, whose Lie algebra is the essentially unique
one-dimensional central extension of the inhomogeneous Galilei
algebra, is given as
\begin{align} \label{eq:Barg}
  \Barg &= \Gal \ltimes (\R^4 \times \Ui),
\end{align}
where the homomorphism $\Gal \to \Aut(\R^4 \times \Ui)$ is given by
\begin{equation} \label{eq:Barg_homom}
  (R,k)(s,y,\e^{\I\varphi}) = (s, Ry + sk,
    \e^{\I(\varphi + k\cdot R y + \frac{1}{2} |k|^2 s)}).
\end{equation}
Note that our sign conventions for the inhomogeneous Galilei group and
the Bargmann group are different from those used in
\cite{Geracie.EtAl:2015}.

\paragraph{Forms on principal bundles}

Forms living on the total space of a principal bundle will be denoted
by boldface letters; their local representatives on the base manifold,
i.e.\ pullbacks along local sections of the bundle, will be denoted by
the corresponding non-boldface letters (the local section being
understood from context).  For example, if $\bomega \in \Omega^1(P,
\mathfrak{g})$ is a connection form on a principal $G$-bundle $P
\xrightarrow{\pi} M$, then for a local section $\sigma \in \Gamma(U,
P)$ on $U \subset M$ the corresponding local connection form is
$\omega = \sigma^*\bomega \in \Omega^1 (U, \mathfrak g)$.

Given a representation $\eta \colon G \to \GL(V)$, the space of
$\eta$-tensorial $k$-forms on the total space $P$ of a principal
$G$-bundle will be denoted by $\Omega^k_\eta (P,V)$ or, if the
representation is clear from context, by $\Omega^k_G(P,V)$.

\paragraph{Index notation}

Unless otherwise specified, spacetime manifolds will be assumed as
four-dimensional and denoted by $M$.  The Einstein summation
convention will be used throughout the article.  Lowercase Greek
letters will be used as coordinate indices on spacetime; for example,
the coordinate component decomposition of a vector field reads $X =
X^\mu \partial_\mu$.

Frame indices labelling a local frame of vector fields, i.e.\ a local
section of the linear frame bundle $F(M)$ or a reduction thereof, will
be uppercase Latin letters.  These will also be used to denote the
frame components of tensor fields; for example, the frame
decomposition of a vector field reads $X = X^A \e_A$, where the local
frame is $(\e_A)$ and the frame components of $X$ are given by $X^A =
\e^A(X) = \e^A_\mu X^\mu$ in terms of the dual frame $(\e^A)$.  Put
differently, denoting by
\begin{equation} \label{eq:assoc_bundle_frame}
  E = F(M) \times_{\GL(4)} \R^4
\end{equation}
the vector bundle associated to the linear frame bundle $F(M)$ via the
fundamental representation of $\GL(4)$, this means that we freely use
the canonical solder form of $E$ to identify $E$ with the tangent
bundle $TM$, while always representing elements of $E$ with respect to
chosen local frames; and analogously for the corresponding tensor
bundles.

If we reduce the structure group of the linear frame bundle to the
Galilei group (for details see section \ref{subsec:Gal_mfs}), and
therefore understand $\R^4$ as the space on which the Galilei group
acts according to \eqref{eq:action_Gal_R4}, we will often decompose
frame indices according to
\begin{subequations}
\begin{equation}
  (A) = (t,a),
\end{equation}
using $t$ as a `temporal' index and lowercase Latin letters as
`spatial' indices running from $1$ to $3$.  For example, a frame of
local vector fields would then be decomposed as
\begin{equation}
  (\e_A) = (\e_t, \e_a),
\end{equation}
or an element of $\R^4$ as
\begin{equation}
  y = (y^A) = (y^t, y^a).
\end{equation}
\end{subequations}
If the relevant group is the \emph{Lorentz} group instead, the
temporal index will be denoted by $0$ instead of $t$.

\paragraph{Linear connections}

Given a linear connection $\nabla$ on a manifold $M$, for its
coordinate connection coefficients $\Gamma^\rho_{\mu \nu}$ the
\emph{first} lower index will be the `form index' / differentiation
index, i.e.
\begin{equation}
  (\nabla_X Y)^\rho
  = X^\mu (\partial_\mu Y^\rho + \Gamma^\rho_{\mu\nu} Y^\nu).
\end{equation}
In particular, the coordinate expression for the torsion is
\begin{equation}
  \tensor{T}{^\rho_{\mu\nu}} = 2 \Gamma^\rho_{[\mu\nu]}
  = \Gamma^\rho_{\mu\nu} - \Gamma^\rho_{\nu\mu} \; .
\end{equation}
Similarly, if we have a `teleparallel' connection $\nabla$ and a
`non-teleparallel' connection $\widetilde\nabla$ (e.g.\ the
Levi-Civita connection in standard Lorentzian \acronym{GR}), for the
components of their difference tensor we will use the convention
\begin{equation} \label{eq:conv_contortion}
  \tensor{K}{^\rho_{\mu\nu}}
  = \Gamma^\rho_{\mu\nu} - \widetilde\Gamma^\rho_{\mu\nu} \; .
\end{equation}
Note that in some literature on teleparallel gravity, in particular in
the review \cite{Bahamonde.EtAl:2021}, the opposite convention is used
for the lower indices of the connection coefficients, with the second
one being the form index.  Therefore, in \cite{Bahamonde.EtAl:2021}
the index positioning on the contortion \eqref{eq:conv_contortion} is
different to ours as well, even though the definition looks identical.

We use the same convention for the coordinate components of local
connection forms with respect to local frames: the form index comes
\emph{before} the frame indices, i.e.\ for a local frame $(\e_A)$ the
local connection form is given by
\begin{equation}
  \nabla \e_B = \tensor{\omega}{^A_B} \otimes \e_A
  \; \text{with} \;
  \tensor{\omega}{^A_B} = \tensor{\omega}{_\mu^A_B} \, \D x^\mu .
\end{equation}

Given a linear connection as a covariant derivative operator $\nabla$
on $TM$, when extending it to higher-degree tensor bundles over $M$
via a Leibniz rule, we will not always employ the identification of
the tangent bundle $TM$ with the associated vector bundle $E$ from
\eqref{eq:assoc_bundle_frame} via the canonical solder form.  We will
extend $\nabla$ to tensor bundles of the form $(TM)^{\otimes r}
\otimes (T^*M)^{\otimes s} \otimes E^{\otimes p} \otimes
(E^*)^{\otimes q}$ in two different ways: one denoted (again) by
$\nabla$, acting \emph{only} on tensor powers of $TM$ or its dual,
\emph{not} acting on the $E$ factors; and another one denoted by $D$,
acting on all factors.  In index notation, this means that covariant
differentiation with $\nabla$ `acts only on spacetime indices', while
covariant differentiation with $D$ `acts on both spacetime and frame
indices'.  For example, this means that for a tensor field
\begin{subequations}
\begin{align}
  X &= X^{\rho A}_{\sigma B} \, \partial_\rho
      \otimes \D x^\sigma \otimes \e_A \otimes \e^B
      \in \Gamma(TM \otimes T^*M \otimes E \otimes E^*) \\
\intertext{we have}
  \nabla_\mu X^{\rho A}_{\sigma B}
  &= \partial_\mu X^{\rho A}_{\sigma B}
    + \Gamma^\rho_{\mu\nu} X^{\nu A}_{\sigma B}
    - \Gamma^\nu_{\mu\sigma} X^{\rho A}_{\nu B} \; , \\
\intertext{but}
  D_\mu X^{\rho A}_{\sigma B}
  &= \partial_\mu X^{\rho A}_{\sigma B}
    + \Gamma^\rho_{\mu\nu} X^{\nu A}_{\sigma B}
    - \Gamma^\nu_{\mu\sigma} X^{\rho A}_{\nu B}
    + \tensor{\omega}{_\mu^A_C} X^{\rho C}_{\sigma B}
    - \tensor{\omega}{_\mu^C_B} X^{\rho A}_{\sigma C} \; .
\end{align}
\end{subequations}

\section{Bargmann structures and teleparallel Galilei connections}
\label{sec:Barg_structs_telep_Gal_conns}

In this section we will introduce the geometric framework for our
teleparallel version of Newton--Cartan gravity, and formulate the
theory.

In subsection \ref{subsec:Gal_mfs}, we review the geometric framework
of Galilei manifolds, which underlies usual Newton--Cartan gravity as
well as its teleparallel variant.  We present the classical
description of the geometry as given by Künzle \cite{Kuenzle:1976} and
Ehlers \cite{Ehlers:1981a,Ehlers:1981b} and reviewed by Malament
\cite[chapter 4]{Malament:2012}, but at some points provide a
modernised perspective based on some aspects of the recent paper
\cite{Geracie.EtAl:2015}.

In subsection \ref{subsec:Barg_structs}, we give a global,
principal-bundle-based formulation of the `Bargmann spacetime'
framework by Geracie \emph{et al.} \cite{Geracie.EtAl:2015}, which
makes explicit in which sense the Bargmann group underlies the
geometry of Galilei maifolds as a local symmetry group.  Of course,
this global description is in some sense implicit in the local
description from~\cite{Geracie.EtAl:2015}; however, an explicitly
global formulation makes the invariant nature of the involved objects
stand out more clearly.

Finally, in subsection \ref{subsec:telep_Gal_conns} we employ the
previously introduced framework to `teleparallelise' Newton--Cartan
gravity.

\subsection{Galilei manifolds}
\label{subsec:Gal_mfs}

A \emph{Galilei manifold} is a 4-dimensional manifold $M$ with a
nowhere vanishing \emph{clock 1-form} $\tau \in \Omega^1(M)$ and a
symmetric contravariant 2-tensor $h$ of signature $(3,0,1)$---i.e.\
positive definite / negative definite / degenerate on subbundles of
$T^*M$ of rank $3$ / $0$ / $1$ respectively---, called the \emph{space
  metric}, whose degenerate direction is spanned by $\tau$, i.e.
\begin{equation}
  \tau_\mu h^{\mu\nu} = 0.
\end{equation}
The kernel of $\tau$ at any point are the \emph{spacelike} vectors at
this point, on which $h$ defines a positive definite scalar product.
The integral of $\tau$ along any curve is interpreted as the time
elapsed along the curve.  In the following, we will mostly assume that
$\D\tau = 0$, such that the time between two events is independent of
the worldline between them chosen to measure it, i.e.\ we have an
absolute notion of time.  This also implies that the distribution of
spacelike vectors is integrable, i.e.\ we have hypersurfaces of
simultaneity / leaves of `space', on each of which $h$ defines a
Riemannian metric.  The assumption $\D\tau = 0$ of absolute time is
made in standard Newton--Cartan gravity.  Note however that by
Frobenius' theorem the integrability of the distribution $\ker\tau$,
i.e.\ the existence of an absolute notion of simultaneity, is
equivalent to the weaker condition $\tau \wedge \D\tau = 0$.
Newton--Cartan gravity in the more general context of a clock form
satisfying only this integrability condition is referred to as
\emph{twistless torsional Newton--Cartan gravity} (\acronym{TTNC}
gravity) \cite{Christensen.EtAl:2014a, Christensen.EtAl:2014b}.  In
this article, we will however only be considering the teleparallel
formulation of standard Newton--Cartan gravity, assuming absolute
time.

A \emph{Galilei frame} is a local frame $(\e_t = v,\e_a)$ of vector
fields on $M$ such that
\begin{equation}
  \tau(v) = 1, \; h^{\mu\nu} = \delta^{ab} \e_a^\mu \e_b^\nu \; .
\end{equation}
A choice of Galilei frame (or, indeed, of a temporal reference vector
field $v$ alone) gives us a projector onto space along $v$, namely
\begin{subequations}
\begin{equation}
  P^\mu_\nu = \delta^\mu_\nu - v^\mu \tau_\nu \; ,
\end{equation}
and defines a $v$-dependent `inverse' of $h$, whose components we
denote by $h_{\mu\nu}$ (always understanding the implicit
$v$-dependence), by
\begin{equation}
  h_{\mu\nu} v^\nu = 0, \;
  h_{\mu\nu} h^{\nu\rho} = P^\rho_\mu \; .
\end{equation}
\end{subequations}
In terms of the vector fields of the frame and the covector fields of
its dual frame, which we denote by $(\e^t = \tau, \e^a)$, these
objects may be expressed as
\begin{equation}
  P^\mu_\nu = \e_a^\mu \e^a_\nu \; , \;
  h_{\mu\nu} = \delta_{ab} \e^a_\mu \e^b_\nu \; .
\end{equation}

In the context of Galilei manifolds, we will use the common convention
of `raising indices' by contraction with the space metric
$h^{\mu\nu}$, and of `lowering' indices by contraction with
$h_{\mu\nu}$ if a temporal reference field $v$ is chosen.  Note that
due to the degeneracy of $h$ these operations are not inverses of each
other, but first raising and then lowering an index (or vice versa)
corresponds to contracting with the spatial projector $P^\mu_\nu$.
Similarly, if a local Galilei frame has been chosen, we will use
$\delta^{ab}$ and $\delta_{ab}$ to raise and lower spatial frame
indices.

As one easily checks, basis change matrices between Galilei frames
(evaluated at a point) are precisely elements of the homogeneous
Galilei group $\Gal$, understood as a subgroup of $\GL(4)$ according
to \eqref{eq:action_Gal_R4}.  Thus, Galilei frames are local sections
of a reduction of the structure group of the linear frame bundle
$F(M)$ from $\GL(4)$ to $\Gal$, which we denote by $G(M)$ and call the
\emph{Galilei frame bundle} of $(M,\tau,h)$.  Given a 4-dimensional
manifold $M$, the specification of such a reduction is equivalent to
the specification of $\tau$ and $h$ making $M$ into a Galilei
manifold, and therefore we call such a reduction $G(M)$ a
\emph{Galilei structure} on $M$.

Any change of local Galilei frame, i.e.\ of local section of
$G(M)$,\footnote{We avoid the ambiguous term `gauge transformation'
  here, since it could mean either such a change of local section or a
  (global) fibre-preserving automorphism of the total space.} has the
form
\begin{subequations} \label{eq:local_Gal_trafo}
\begin{equation}
  (v, \e_a) \to (v, \e_a) \cdot (R,k)^{-1}
\end{equation}
for a local $\Gal$-valued function $(R,k)$, where the dot denotes the
right action of $\Gal$ on $G(M)$.  Spelled out, this reads
\begin{equation}
  (v, \e_a) \to \left(v - \e_b \tensor{(R^{-1})}{^b_a} k^a,
    \e_b \tensor{(R^{-1})}{^b_a}\right) ,
\end{equation}
\end{subequations}
and we call such a change of frame a \emph{local Galilei
  transformation} by $(R,k)$.  This defines a left action on local
Galilei frames by the group of local $\Gal$-valued functions.

A \emph{Galilei connection} on a Galilei manifold is a principal
connection $\bomega$ on the Galilei frame bundle $G(M)$.  The local
connection form with respect to a Galilei frame $(v,\e_a)$, i.e.\ the
pullback of the connection form $\bomega$ on the total space along the
frame, is a local one-form $\omega$ on $M$ with values in the
homogeneous Galilei Lie algebra $\gal = \so(3) \oright \R^3$.  We
decompose it as
\begin{equation}
  \omega = (\tensor{\omega}{^a_b}, \varpi^a),
\end{equation}
and following \cite{Geracie.EtAl:2015}, we will sometimes call its
$\so(3)$-valued part $\tensor{\omega}{^a_b}$ the \emph{spin
  connection} and its $\R^3$-valued part $\varpi^a$ the \emph{boost
  connection}.\footnote{Note that due to the semidirect product nature
  of $\Gal$, the boost `connection' does not define a connection by
  itself, other than the spin connection, which induces a connection
  on the bundle of orthonormal frames for the distribution $\ker\tau
  \subset TM$ of spacelike vectors together with the metric induced by
  $h$.}

Since $G(M)$ is a reduction of the structure group of $F(M)$ to
$\Gal$, a Galilei connection is equivalently given by a covariant
derivative operator $\nabla$ on the tangent bundle $TM$ whose local
connection form $\tensor {\omega}{^A_B}$ with respect to a Galilei
frame, defined by
\begin{equation}
  \nabla\e_B = \tensor{\omega}{^A_B} \otimes \e_A
\end{equation}
and a priori taking values in $\gl(4)$, takes values in $\gal$, viewed
as a subalgebra of $\gl(4)$ according to \eqref{eq:action_Gal_R4}.
This means that $\tensor{\omega} {^A_B}$ satisfies
\begin{subequations} \label{eq:connection_gal}
\begin{align}
  \tensor{\omega}{^t_A} &= 0, \\
  \delta_{ac} \tensor{\omega}{^c_b}
    &= - \delta_{bc} \tensor{\omega}{^c_a} \; .
\end{align}
\end{subequations}
The boost connection $\varpi^a$ then arises as the `spatio-temporal
part' of $\tensor{\omega}{^A_B}$,
\begin{equation}
  \varpi^a = \tensor{\omega}{^a_t} \; .
\end{equation}
Equivalently to \eqref{eq:connection_gal}, a covariant derivative
operator $\nabla$ is a Galilei connection iff it is compatible with
$\tau$ and $h$, i.e.\ satisfies
\begin{equation}
  \nabla \tau = 0, \; \nabla h = 0.
\end{equation}
Note that from compatibility with $\tau = \e^t$, it follows that for
any tensor field with components $X^{\mu_1 \dots \mu_{r+1}}_{\nu_1
  \dots \nu_s}$ satisfying $X^{\mu_1 \dots \mu_r t}_{\nu_1 \dots
  \nu_s} = 0$ we have
\begin{equation} \label{eq:cov_deriv_t_zero}
  D_\sigma X^{\mu_1 \dots \mu_r t}_{\nu_1 \dots \nu_s} = 0.
\end{equation}

The \emph{Newton--Coriolis form}\footnote{This name was introduced in
  \cite{Geracie.EtAl:2015}, after previous authors had either given
  the form no name or called it just the ‘Coriolis form’.} of a
Galilei connection $\bomega$ with respect to a local Galilei frame
$(v,\e_a)$ is the (local) two-form $\Omega$ on $M$ with components
\begin{subequations}
\begin{align}
  \Omega_{\mu\nu} &:= 2 (\nabla_{[\mu}v^\kappa) h_{\nu]\kappa} \; . \\
\intertext{Due to $\nabla v = \nabla \e_t = \varpi^a \otimes \e_a$, we
  may also express $\Omega$ in terms of the boost connection,}
  \Omega_{\mu\nu} &= 2 \tensor{\varpi}{_{[\mu}^a}
    \tensor{h}{_{\nu]\kappa}} \e_a^\kappa
  = 2 \varpi_{[\mu|a|} \e^a_{\smash{\nu]}} \; , \\
\intertext{which may also be written as}
  \Omega &= \varpi_a \wedge \e^a . \\
\intertext{Using the notation $\tensor{\varpi}{_\mu^\nu} :=
  \tensor{\varpi}{_\mu^a} \e_a^{\nu}$, this may be expressed by the
  mnemonic that `$\Omega$ is the antisymmetric part of the boost
  connection',}
  \Omega_{\mu\nu} &= 2 \varpi_{[\mu\nu]} \; .
\end{align}
\end{subequations}
However, this involves some abuse of notation: since
$\tensor{\varpi}{_\mu^a}$ are not the components of a tensor field on
$M$, the components $\tensor{\varpi} {_\mu^\nu}$ don't have an
invariant meaning.

A Galilei connection is determined by its torsion $T$ and its
Newton--Coriolis form $\Omega$, with connection coefficients in
coordinates given by
\begin{equation} \label{eq:Galilei_conn_classif}
  \Gamma^\sigma_{\mu\nu}
  = v^\sigma \partial_{(\mu} \tau_{\nu)}
  + \frac{1}{2} h^{\sigma\rho} (\partial_\mu h_{\nu\rho}
    + \partial_\nu h_{\mu\rho} - \partial_\rho h_{\mu\nu})
  + \frac{1}{2} \tensor{T}{^\sigma_{\mu\nu}}
  - \tensor{T}{_{(\mu\nu)}^\sigma}
  + \tensor{\tau}{_{(\mu}} \tensor{\Omega}{_{\nu)}^\sigma} .
\end{equation}
The temporal component of the torsion of a Galilei connection is equal
to $\tensor{T}{^t_{\mu\nu}} = (\D\tau)_{\mu\nu}$, i.e.\ if we assume
absolute time, the temporal torsion vanishes.  If conversely $\Omega$
is an arbitrary two-form and $T$ any tensor field with the
antisymmetry of a torsion that satisfies $\tensor{T}{^t_{\mu\nu}} =
(\D\tau)_{\mu\nu}$, then \eqref{eq:Galilei_conn_classif} defines a
Galilei connection with torsion $T$ and Newton--Coriolis form
$\Omega$.

We note that, as for any principal connection on a reduction of the
linear frame bundle, the torsion of a Galilei connection can be
written in the following way: the tangent bundle $TM$ is canonically
isomorphic to the associated vector bundle $E = G(M) \times_\Gal \R^4$
via the canonical solder form $\theta \in \Omega^1(M,E)$ of $E$.
Taking the exterior covariant derivative of the solder form with
respect to the connection then yields the torsion $T \in \Omega^2(M,E)
\cong \Omega^2(M,TM)$ (`Cartan's first structure equation'):
\begin{subequations} \label{eq:structure_first}
\begin{equation}
  T = \D^\bomega \theta
\end{equation}
In terms of a local Galilei frame, we have $\theta = \e^A \otimes
\e_A$, i.e.\ the components of the canonical solder form with respect
to the frame (viewed as a local frame for $E$) are given by the dual
frame one-forms.  Thus, expressed in components the first structure
equation reads
\begin{equation}
  T^A = (\D^\bomega \theta)^A
  = \D\e^A + \tensor{\omega}{^A_B} \wedge \e^B ,
\end{equation}
or explicitly for Galilei connections and Galilei frames
\begin{equation}
  (T^t, T^a) = (\D\tau, \D e^a
    + \tensor{\omega}{^a_b} \wedge \e^b + \varpi^a \wedge \tau).
\end{equation}
\end{subequations}

A Galilei connection on a Galilei manifold with absolute time ($\D\tau
= 0$) is called \emph{Newtonian} iff it is torsion-free and its
curvature tensor satisfies symmetry in pairs,
\begin{subequations}
\begin{equation}
  \tensor{R}{^\mu_\nu^\rho_\sigma}
  = \tensor{R}{^\rho_\sigma^\mu_\nu} \; .
\end{equation}
This is equivalent (which is not at all obvious!) to the
Newton--Coriolis form (with respect to any frame) being closed,
\begin{equation}
  \D\Omega = 0.
\end{equation}
\end{subequations}
At each point, the curvature tensor of a Newtonian connection has as
many linearly independent components as that of the Levi-Civita
connection of a (pseudo-)Riemannian metric on a manifold of the same
dimension, i.e.\ 20 in the case of four spacetime dimensions.

We now consider how the objects introduced above transform under local
Galilei transformations, i.e.\ under changes of Galilei frame
parametrised by local $\Gal$-valued functions $(R,k)$ according to
\eqref{eq:local_Gal_trafo}.  To keep the formulae easier, we consider
the transformation under local frame rotations and under local Galilei
boosts separately.  Spelled out, the purely rotational transformation
of the frame with parameter $R$ reads
\begin{subequations} \label{eq:local_rot}
\begin{equation}
  (v, \e_a) \to \left(v, \e_b \tensor{(R^{-1})}{^b_a}\right) ,
\end{equation}
and those of the dual frame and the connection form are
\begin{align}
  (\tau, \e^a) &\to (\tau, \tensor{R}{^a_b} \e^b) , \\
  \tensor{\omega}{^a_b} &\to \tensor{R}{^a_c}
      \tensor{\omega}{^c_d} \tensor{(R^{-1})}{^d_b}
    + \tensor{R}{^a_c} \D\tensor{(R^{-1})}{^c_b} \; , \\
  \varpi^a &\to \tensor{R}{^a_b}\varpi^b ,
\end{align}
\end{subequations}
while for pure boosts with parameter $k$ (which sometimes are called
\emph{Milne boosts} instead of `local Galilei boosts'), we have
\begin{subequations} \label{eq:local_boost}
\begin{align}
  (v, \e_a) &\to \left(v - \e_a k^a, \e_a\right) , \\
  (\tau, \e^a) &\to (\tau, \e^a + k^a \tau) , \\
  \tensor{\omega}{^a_b} &\to \tensor{\omega}{^a_b} \; , \\
  \varpi^a &\to \varpi^a - \D k^a - \tensor{\omega}{^a_b} k^b .
\end{align}
\end{subequations}

\subsection{Bargmann structures}
\label{subsec:Barg_structs}

Given a Galilei manifold $(M,\tau,h)$, we can extend its Galilei frame
bundle $G(M)$ to a principal bundle with structure group the Bargmann
group $\Barg = \Gal \ltimes (\R^4 \times \Ui)$ in the following
way.  We denote by
\begin{subequations}
\begin{align}
  \rho \colon \Gal &\to \Aut(\R^4 \times \Ui) \\
\intertext{the group homomorphism defining the Bargmann group, given
  in \eqref{eq:Barg_homom}; by}
  \dot\rho \colon \Gal &\to \Aut(\R^4 \oplus \ui) \\
\intertext{the induced representation of $\Gal$ by Lie algebra
  automorphisms, which explicitly takes the form}
  \label{eq:rhodot_explicit}
  \dot\rho_{(R,k)} (y^t, y^a, \I\varphi)
  &= \left( y^t, \tensor{R}{^a_b} y^b + y^t k^a,
    \I(\varphi + \tfrac{1}{2} |k|^2 y^t
      + k_a \tensor{R}{^a_b} y^b) \right) \\
\intertext{for $(R,k) \in \Gal$ and $(y^A, \I\varphi) \in \R^4 \oplus
  \ui$; and by}
  \tilde\rho \colon \Gal &\to \Diff(\Barg)
\end{align}
\end{subequations}
the left action of $\Gal$ on $\Barg$ by multiplication.  The
associated bundle
\begin{equation}
  B(M) := G(M) \times_{\tilde\rho} \Barg
\end{equation}
is then a principal $\Barg$-bundle (with action given by right
multiplication) that extends $G(M)$ along the natural inclusion map
\begin{equation}
  \gamma \colon G(M) \to B(M), \gamma(p) = [p, \e_\Barg].
\end{equation}
(Here $\e_\Barg$ is the neutral element of $\Barg$ and we denoted the
element of the Galilei frame bundle by $p$ instead of our usual
notation $(\e_A)$ in order to avoid confusion with the neutral
element.)

The fundamental observation is now that connections $\bhatomega$ on
this bundle $B(M)$ are in one-to-one correspondence with pairs
$(\bomega, \bTheta)$ of connections $\bomega$ and $\dot\rho$-tensorial
one-forms $\bTheta \in \Omega^1_{\dot\rho}(G(M), \R^4 \oplus \ui)$ on
$G(M)$ via the pullback condition
\begin{subequations}
\begin{align}
  \gamma^*\bhatomega &= (\bomega,\bTheta). \\
\intertext{Furthermore, in this situation the pullback of the
  curvature form $\boldsymbol{\hat R}$ of $\bhatomega$ is given by the
  curvature form $\boldsymbol{R}$ of $\bomega$ and the exterior
  covariant derivative of $\bTheta$ with respect to $\bomega$,}
  \label{eq:Bargmann_conn_curv}
  \gamma^*\boldsymbol{\hat R}
  &= (\boldsymbol{R}, \D^\bomega\bTheta).
\end{align}
\end{subequations}
This generalises the classical situation of connections on the affine
frame bundle of a manifold (see, e.g., \cite [section
III.3]{Kobayashi.Nomizu:1963}), and is generally true for `semidirect
extensions' of principal bundles as encountered here.  Details and a
general discussion of this extension construction may be found in
appendix \ref{sec:appendix_semidir_extension}.

We further decompose
\begin{equation}
  \bTheta = (\btheta, \I\ba)
\end{equation}
with $\btheta \in \Omega^1(G(M), \R^4)$ and $\ba \in \Omega^1(G(M))$.
Due to $\ui \subset \R^4 \oplus \ui$ being a $\dot\rho$-invariant
subspace, we may view the $\R^4$-valued part $\btheta$ as transforming
under the quotient representation $\Gal \to \GL((\R^4\oplus\ui)/\ui)
\cong \GL(4)$, which is the usual representation \eqref
{eq:action_Gal_R4} of $\Gal$ on $\R^4$.  This means that $\btheta$ is
by itself a tensorial form
\begin{equation}
  \btheta \in \Omega^1_\Gal(G(M), \R^4),
\end{equation}
which naturally corresponds to an associated-bundle-valued form
\begin{equation}
  \theta \in \Omega^1(M, G(M) \times_\Gal \R^4)
\end{equation}
on our Galilei manifold.  If this is the canonical solder form of
$G(M) \times_\Gal \R^4$, then we call $\ba \in \Omega^1(G(M))$ a
\emph{Bargmann structure} on $(M,\tau,h)$.\footnote{Note that in
  \cite{Duval.EtAl:1985}, the term `Bargmann structure' was used
  differently, namely for the structure of a five-dimensional
  Lorentzian manifold with a null isometry from which a Galilei
  manifold may be obtained by null reduction; see also
  \cite{Julia.Nicolai:1995}.}

Summed up, a Bargmann structure on a Galilei manifold $(M,\tau,h)$ is
a one-form $\ba \in \Omega^1(G(M))$ on the Galilei frame bundle that
together with the tensorial form $\btheta \in \Omega^1_\Gal(G(M),
\R^4)$ corresponding to the canonical solder form combines into a
$\dot\rho$-tensorial form $(\btheta, \I\ba) \in \Omega^1_{\dot\rho}
(G(M), \R^4 \oplus \ui)$, which in turn together with a Galilei
connection $\bomega \in \Omega^1(G(M), \gal)$ would give a `Bargmann
connection' $\bhatomega$ on $B(M)$.  Note however that we consider the
choice of Galilei connection $\bomega$ \emph{not} to be part of the
choice of Bargmann structure.

Given a Bargmann structure $\ba$, the tensorial form $\bTheta =
(\btheta, \I\ba) \in \Omega^1_{\dot\rho}(G(M), \R^4 \oplus \ui)$
corresponds to an associated-bundle-valued form $\Theta \in
\Omega^1(M, G(M) \times_{\dot\rho} (\R^4 \oplus \ui))$.  The local
representative of this form with respect to a local Galilei frame
$\sigma = (\e_A)$ defined on an open set $U \subset M$, i.e.\ the
pullback
\begin{subequations}
\begin{align}
  (\tau, \e^a, \I a) = \sigma^* \bTheta
    &\in \Omega^1(U, \R^4 \oplus \ui) \\
\intertext{along the frame, using which we can locally express
  $\Theta$ as}
  \Theta = [\sigma, (\tau, \e^a, \I a)]
    &\in \Omega^1\left( U, G(M) \times_{\dot\rho}
      (\R^4 \oplus \ui) \right),
\end{align}
\end{subequations}
is what in \cite{Geracie.EtAl:2015} was called an \emph {extended
  coframe}.  (Note that since $\btheta$ corresponds to the canonical
solder form, its components with respect to the local frame $\sigma =
(\e_A) = (v, \e_a)$ are given by the dual frame $(\tau, \e^a)$.)

Under local Galilei transformations, i.e.\ changes of local Galilei
frame, the extended coframe transforms according to the representation
$\dot\rho$.  Using the explicit form \eqref{eq:rhodot_explicit} of
$\dot\rho$, we see that this means that while the dual frame $(\tau,
\e^a)$ transforms as in \eqref{eq:local_rot}, \eqref{eq:local_boost},
the one-form $a$ locally representing the Bargmann structure is
invariant under local rotations, and under local boosts with parameter
$k$ transforms as
\begin{equation} \label{eq:local_boost_Bargmann_str}
  a \to a + k_a \e^a + \frac{1}{2} |k|^2 \tau .
\end{equation}

We now consider the exterior covariant derivative $\D ^\bomega\Theta$
of the form $\Theta$, locally represented by the extended coframe,
with respect to a Galilei connection $\bomega$.  According to
\eqref{eq:Bargmann_conn_curv}, it corresponds to the $\R^4 \oplus
\ui$-valued part of the curvature of the `Bargmann connection'
$\bhatomega$ on $B(M)$ given by $\bomega$ and $\bTheta$.  Since the
$\R^4$-valued part of $\bTheta$ corresponds to the canonical solder
form of $G(M) \times_\Gal \R^4$, according to Cartan's first structure
equation \eqref{eq:structure_first} the $\R^4$-valued part of
$\D^\bomega \bTheta$ corresponds to the torsion of $\bomega$.
Following \cite{Geracie.EtAl:2015}, we will call $\D^\bomega\Theta$
the connection's \emph{extended torsion} with respect to the Bargmann
structure, and denote its local components with respect to a local
frame $\sigma = (\e_A)$ by
\begin{equation}
  (T^A, \I f) := \sigma^*(\D^\bomega\bTheta).
\end{equation}
Its $\ui$-valued part $f$, which of course does not define an
invariant geometric object on its own, we call the \emph {mass
  torsion} following \cite{Geracie.EtAl:2015}.  From the explicit form
\eqref{eq:rhodot_explicit} of $\dot\rho$, we obtain the induced Lie
algebra representation $\dot\rho' \colon \gal \to \Der(\R^4 \oplus
\ui)$ as
\begin{equation}
  \dot\rho'_{(X,k)}(y^A, \I\varphi)
  = \left( ((X,k)y)^A, \I k_a y^a \right)
\end{equation}
for $(X,k) \in \gal = \so(3) \oright \R^3$ and $(y^A, \I\varphi) \in
\R^4 \oplus \ui$.  Therefore, the local components of the extended
torsion are\footnote{Under local Galilei transformations, the extended
  torsion of course again transforms according to $\dot\rho$, i.e.\
  under local rotations the mass torsion is invariant, and under local
  Galilei boosts with parameter $k$ it transforms as
  \begin{equation}
    f \to f + k_a T^a + \frac{1}{2} |k|^2 \D\tau.
  \end{equation}
  However, we won't need this in our subsequent discussion.}
\begin{subequations}
\begin{align}
  (T^A, \I f) &= \sigma^*(\D^\bomega \bTheta) \nonumber\\
              &= \D(\e^A, \I a) + \dot\rho'_{(\omega,\varpi)}
                \wedge (\e^A, \I a) \nonumber\\
              &= (\D\e^A + \tensor{\omega}{^A_B} \wedge \e^B,
                \I (\D a + \varpi_a \wedge \e^a)),
\end{align}
and the mass torsion is given by
\begin{equation}
  f = \D a + \varpi_a \wedge \e^a .
\end{equation}
\end{subequations}

In terms of the mass torsion $f$ and the $\ui$ component $a$ of the
extended coframe that locally represents the Bargmann structure, we
can write the Newton--Coriolis form of the connection as
\begin{equation}
  \Omega = \varpi_a \wedge \e^a = f - \D a ,
\end{equation}
i.e.\ given a choice of Bargmann structure, $\Omega$ is determined by
(and determines) the mass torsion $f$.  Since the freedom in the
choice of a Galilei connection lies precisely in the torsion and the
Newton--Coriolis form, we thus see that on a Galilei manifold with a
Bargmann structure, \emph{a Galilei connection is uniquely
  characterised by its extended torsion}.  In particular, if we have
absolute time (i.e.\ $\D\tau = 0$), which we will from now on assume
for all Galilei manifolds unless otherwise stated, for each Bargmann
structure there is a unique Galilei connection with vanishing extended
torsion.  Since its Newton--Coriolis form is closed (it's even
exact!), this unique extended-torsion-free connection is Newtonian.

Note that we have a $\Ui$ gauge freedom for Bargmann structures,
corresponding to the $\Ui$ direction in the Bargmann group: given a
Bargmann structure $\ba$ on $(M,\tau,h)$ and any $\Ui$-valued function
$\e^{\I\chi}$ on $M$, we may act with it (actively!) on $\ba$ and
obtain a \emph{new} Bargmann structure\footnote{From the global point
  of view, this transformation arises as follows.  From $\e^{\I\chi}$,
  we obtain a gauge transformation $f_\chi$ of the principal
  $\Barg$-bundle $B(M) \overset{\hat\pi}{\to} M$, i.e.\ a
  fibre-preserving principal bundle automorphism $f_\chi \colon B(M)
  \to B(M)$, as $f_\chi(p) := p \cdot (\mathbb{1}, 0, 0, \e^{\I\chi
    \circ \hat\pi})$.  Given a connection $\bhatomega$ on $B(M)$, we
  may act on it with the gauge transformation $f_\chi$, giving the new
  connection $f_\chi^* \bhatomega = \bhatomega + (0, 0, 0, \I
  \hat\pi^*(\D\chi))$.  For the case of $\bhatomega$ given by a
  Galilei connection $\bomega$ and a Bargmann structure $\ba$, this
  gives rise to \eqref{eq:Barg_Ui_trafo}.}
\begin{subequations} \label{eq:Barg_Ui_trafo}
\begin{align}
  \ba &\to \ba + \pi^*(\D\chi), \\
  \intertext{which on $M$ is locally represented by}
  a &\to a + \D\chi .
\end{align}
\end{subequations}
Note that with respect to two Bargmann structures which are related to
each other by such a $\Ui$ gauge transformation, a Galilei connection
has the same extended torsion, since $f = \D a + \Omega = \D(a +
\D\chi) + \Omega$.

\subsection{Teleparallel Galilei connections}
\label{subsec:telep_Gal_conns}

In Lorentzian (or more generally pseudo-Riemannian) geometry, any
metric connection is uniquely determined by its (arbitrarily
specifiable) torsion.  Therefore, the difference tensor between an
arbitrary metric connection $\nabla$ and the torsion-free Levi-Civita
connection may be expressed purely in terms of the torsion of
$\nabla$.  This allows for a reformulation of the Einstein equation,
which is usually formulated in terms of the Levi-Civita connection, in
terms of a flat torsionful connection $\nabla$ and its torsion, giving
rise to \acronym{TEGR}, the teleparallel equivalent of general
relativity.

Differently to that, in the classical setting of Newton--Cartan
gravity, such a reformulation is not possible, since Galilei
connections on a Galilei manifold are not uniquely determined by their
torsion: according to \eqref{eq:Galilei_conn_classif}, we also need to
specify the Newton--Coriolis form $\Omega$.  Since $\Omega$ depends on
the choice of timelike Galilei frame vector field $v$, there is no
naturally given unique connection $\widetilde\nabla$ the difference to
which of a general connection $\nabla$ we could use to reformulate
Newton--Cartan gravity.

However, this is remedied by the introduction of a Bargmann structure.
As we have seen above, on a Galilei manifold with absolute time with a
chosen Bargmann structure, there is a unique Galilei connection with
vanishing extended torsion.  This allows us to reformulate
Newton--Cartan gravity in a teleparallel way, which we will explain in
the following.

\newcommand{\Ksp}{\accentset{(3)}{K}}

Given a Galilei manifold $(M,\tau,h)$ with absolute time (i.e.\
$\D\tau = 0$) with a Bargmann structure $\ba$, we define the
\emph{Newton--Cartan contortion} of a Galilei connection $\bomega$ to
be the tensor field
\begin{subequations} \label{eq:defn_NC_contort}
\begin{equation}
  \tensor{K}{^\rho_{\mu\nu}}
  := \tensor{\Ksp}{^\rho_{\mu\nu}}
  + \tensor{\tau}{_{\smash{(\mu}}} \tensor{f}{_{\smash{\nu)}}^\rho}
\end{equation}
with
\begin{equation}
  \tensor{\Ksp}{^\rho_{\mu\nu}}
  := \frac{1}{2} \tensor{T}{^\rho_{\mu\nu}}
  - \tensor{T}{_{(\mu\nu)}^\rho} ,
\end{equation}
\end{subequations}
where $(T^A, \I f)$ are the local components of the extended torsion
of $\bomega$ with respect to any local Galilei frame.  Comparing this
to the general form of the coordinate connection coefficients of a
Galilei connection \eqref{eq:Galilei_conn_classif}, we see that
$\tensor{K} {^\rho_{\mu\nu}}$ is the difference tensor between our
connection $\bomega$ and the unique extended-torsion-free connection
$\boldsymbol{\widetilde\omega}$, i.e.
\begin{equation}
  \tensor{K}{^\rho_{\mu\nu}}
  = \Gamma^\rho_{\mu\nu}
  - \widetilde\Gamma^\rho_{\mu\nu} \; .
\end{equation}
Note that this also implies that despite being defined in terms of
objects depending on the choice of local Galilei frame in
\eqref{eq:defn_NC_contort}, the Newton--Cartan contortion is in fact
independent of the choice of frame.\footnote{This may of course also
  be verified by direct calculation, using the transformation
  behaviour of the mass torsion and the local coframe (which enters
  the definition of $\tensor {K}{^\rho_{\mu\nu}}$ through lowering of
  indices with $h_{\mu\nu}$).}

Using the notion of Newton--Cartan contortion, we can now formulate
teleparallel Newton--Cartan gravity in terms of the following axioms:

\paragraph{Axioms for teleparallel Newton--Cartan gravity}

\begin{enumerate}[label=(\roman*)]
\item Spacetime is a Galilei manifold $(M,\tau,h)$ with absolute time,
  endowed with a Bargmann structure and a flat Galilei connection
  $\bomega$,
\item ideal clocks measure time as defined by $\tau$, and ideal rods
  measure spatial lengths as defined by the metric induced by $h$ on
  spacelike vectors,
\item free test particles move on timelike curves $\gamma$ solving
  \begin{equation} \label{eq:EoM_telNC}
    (\nabla_{\dot\gamma} \dot\gamma)^\rho
    = \tensor{K}{^\rho_{\mu\nu}} \dot\gamma^\mu \dot\gamma^\nu ,
  \end{equation}
\item the field equation
  \begin{equation} \label{eq:field_eq_telNC}
    - \tensor{D}{_\sigma} \tensor{K}{^\sigma_{AB}}
    + \tensor{D}{_A} \tensor{K}{^\mu_{\mu B}}
    - \tensor{K}{^\mu_{\sigma B}} \tensor{T}{^\sigma_{\mu A}}
    + \tensor{K}{^\mu_{\mu\sigma}} \tensor{K}{^\sigma_{AB}}
    - \tensor{K}{^\mu_{A\sigma}} \tensor{K}{^\sigma_{\mu B}}
    = 4\pi G \, \rho \, \tensor{\tau}{_A} \tensor{\tau}{_B}
  \end{equation}
  holds, where $\rho$ is the mass density.
\end{enumerate}
The theory defined by these axioms will be our subject of study in the
rest of this article.  Note that the left-hand side of the field
equation \eqref{eq:field_eq_telNC} are just the components $\widetilde
R_{AB}$ of the Ricci tensor of the extended-torsion-free connection
$\boldsymbol{\widetilde\omega}$, expressed in terms of the connection
$\bomega$, its torsion and the difference tensor between the two
connections (i.e.\ the Newton--Cartan contortion of $\bomega$).
Therefore, the above theory really is an equivalent formulation of
usual Newton--Cartan gravity, whose equation of motion and field
equation read $\widetilde \nabla_{\dot\gamma}\dot\gamma = 0$ and
$\widetilde R_{\mu \nu} = 4\pi G \, \rho \, \tau_\mu \tau_\nu$,
respectively.

Note that the formulation of teleparallel Newton--Cartan gravity given
here is completely general, i.e.\ prior to any `gauge-fixing' of
either the connection or the frame.  This distinguishes our
formulation from that given in \cite{Read.Teh:2018}, where both the
field equation and the test particle equation of motion were only
constructed from those of usual Newton--Cartan gravity in a
gauge-fixed situation.

\section{Teleparallel Newton--Cartan gravity from \acronym{TEGR}}
\label{sec:telepar_NC_from_TEGR}

In this section, we will show that teleparallel Newton--Cartan gravity
as introduced in the previous section arises as the formal $c \to
\infty$ limit of \acronym{TEGR}, $c$ being the speed of light, in
direct analogy to standard Newton--Cartan gravity being the formal $c
\to \infty$ limit of general relativity.

\newcommand{\TL}{\accentset{\text{L}}{T}}
\newcommand{\KL}{\accentset{\text{L}}{K}}
\newcommand{\nablaL}{\accentset{\text{L}}{\nabla}}
\newcommand{\DL}{\accentset{\text{L}}{D}}
\newcommand{\omegaL}{\accentset{\text{L}}{\omega}}
\newcommand{\bomegaL}{\accentset{\text{L}}{\boldsymbol{\omega}}}
\newcommand{\GammaL}{\accentset{\text{L}}{\Gamma}}
\newcommand{\nablatildeL}{\accentset{\text{L}}{\widetilde\nabla}}
\newcommand{\GammatildeL}{\accentset{\text{L}}{\widetilde\Gamma}}
\newcommand{\RL}{\accentset{\text{L}}{R}}

\subsection{Formal expansions of Lorentzian geometry}

In the following, we will describe how the formal $c \to \infty$ limit
of a Lorentzian manifold gives rise to a Galilei manifold with a
Bargmann structure, and how in this limit a Lorentzian metric
connection becomes a Galilei connection.  Most of the claims are not
immediately obvious, but all of them may be verified by direct
calculation.

In order to perform the formal $c \to \infty$ limit, we will expand
all objects of Lorentzian geometry as formal power series in the
parameter $c^{-1}$---or, more precisely, formal Laurent series, since
we will need negative orders of $c^{-1}$.  The `formal $c \to \infty$
limit' of a quantity expanded as a power series will then be the term
of order $\Or(c^0)$, provided that there are no terms of negative
order in $c^{-1}$.

Of course, analytically speaking, a `Taylor expansion' in a
dimensionful parameter like $c$ does not make sense (even more so
since $c$ is a constant of nature); only for \emph{dimensionless}
parameters can a meaningful `small-parameter approximation' be made.
In physical realisations of the limit from Lorentzian to Galilei
geometry, this means that the corresponding small parameter has to be
chosen as, e.g., the ratio of some typical velocity of the system
under consideration to the speed of light.  A rigorous discussion of
the Newtonian limit from standard \acronym{GR} to Newton--Cartan
gravity in terms of an actual small parameter approaching zero is
given by Ehlers in his article on `frame theory'
\cite{Ehlers:1981a,Ehlers:1981b}; a nice discussion on the
relationship of formal Newtonian limits to actual physical
slow-velocity approximations may be found in \cite[section II]
{Tichy.Flanagan:2011}.

In the following, however, we will ignore these issues and just expand
in $c^{-1}$ as a formal parameter, thus only considering how the
Lorentzian theory may be viewed as a deformation of its formal
`Newtonian limit'.  Our formal expansion in $c^{-1}$ to implement the
Newtonian limit is the same as that used in \cite{Hansen.EtAl:2019,
  Hansen.EtAl:2020} for the geometric description of post-Newtonian
expansions of \acronym{GR}, specialised to the case $\D\tau = 0$ of
absolute time, but at the same time generalised to allow for
torsionful Lorentzian connections.

To notationally distinguish Lorentzian geometric objects from their
Newton--Cartan counterparts, we will mostly denote the Lorentzian
objects by an overset `L'; for example, the torsion of a Lorentzian
connection will be denoted by $\TL$.

We start with a Lorentzian metric $g$ on our spacetime manifold $M$,
for which we have a local orthonormal frame / tetrad $(\E_A)$ with
dual frame $(\E^A)$, such that the metric and inverse metric can be
written as
\begin{equation}
  g = \eta_{AB} \E^A \otimes \E^B, \quad
  g^{-1} = \eta^{AB} \E_A \otimes \E_B ,
\end{equation}
where $\eta_{AB}$ denotes the components of the Minkowski metric in
Lorentzian coordinates, i.e.\ $(\eta_{AB}) = \mathrm{diag}
(-1,1,1,1)$.  We assume that the frame and dual frame may be expanded
as formal power series in $c^{-1}$ as
\begin{subequations} \label{eq:Lor_expansion_tetr}
\begin{align}
  \E^0_\mu &= c \tau_\mu + c^{-1} a_\mu + \Or(c^{-3}),
  & \E^a_\mu &= \e^a_\mu + \Or(c^{-2}), \\
  \E_0^\mu &= c^{-1} v^\mu + \Or(c^{-3}),
  & \E_a^\mu &= \e_a^\mu + \Or(c^{-2})
\end{align}
\end{subequations}
for some nowhere vanishing one-form $\tau$.  From these assumptions it
follows that $\tau$ and $h := \delta^{ab} \e_a \otimes \e_b$ make $M$
into a Galilei manifold.  For this Galilei manifold that arises as the
formal $c \to \infty$ limit of our Lorentzian manifold, $(v,\e_a)$ is
a local Galilei frame with dual frame $(\tau, \e^a)$.  Let us stress
that $\tau$ is to be viewed as an `input' for the expansion; it is the
object with respect to which we perform the formal Newtonian limit.

We now consider a local Lorentz boost $(\tensor{\Lambda}{^A_B})$
parametrised by the $\R^3$-valued boost velocity function $k$
as\footnote{Note that this arises from expanding the standard form
  $\Lambda = \exp(\zeta^a K_a)$ of a boost, written in terms of the
  rapidity $\zeta^a = \mathrm{artanh}(|k|/c) \frac{k^a}{|k|}$ and the
  boost generators $\tensor{(K_a)}{^A_B} = \delta^A_0 \eta_{aB} -
  \delta^A_a \eta_{0B}$.}
\begin{subequations}
\begin{align}
  \tensor{\Lambda}{^0_0} &= 1 + c^{-2} \frac{|k|^2}{2} + \Or(c^{-4}), \\
  \tensor{\Lambda}{^a_0} &= c^{-1} k^a + \Or(c^{-3})
    = \delta^{ab} \tensor{\Lambda}{^0_b} \; , \\
  \tensor{\Lambda}{^a_b} &= \delta^a_b + c^{-2} \frac{k^a k_b}{2}
    + \Or(c^{-4}).
\end{align}
\end{subequations}
Transforming the Lorentzian frame $(\E_A)$ by $\Lambda$ according to
\begin{equation}
  (\E_A) \to (\tilde\E_A)
  = \left( \E_B \tensor{(\Lambda^{-1})}{^B_A} \right),
\end{equation}
and expanding the new frame analogously to
\eqref{eq:Lor_expansion_tetr}, we obtain a local Galilei boost of the
Galilei frame $(v,\e_a)$ with boost velocity parameter $k$.
Furthermore, under this change of frame, the local one-forms $a$ that
arise as the $c^{-1}$ component of the timelike dual frame one-form
$\E^0$ transform according to \eqref{eq:local_boost_Bargmann_str},
thereby defining a Bargmann structure $\ba$ on $(M,\tau,h)$.

This means that \emph{as the formal $c \to \infty$ limit of the
  Lorentzian manifold we started with, we obtain a Galilei manifold
  with a Bargmann structure}.  We stress again that the only
assumption that is needed for this result is an expansion of the
Lorentzian tetrad and dual tetrad as in \eqref{eq:Lor_expansion_tetr},
with a nowhere vanishing $\tau$.

On the Lorentzian manifold $(M,g)$, we now also consider a metric
connection $\bomegaL$ which we assume to have a regular formal $c \to
\infty$ limit, by which we mean that its coordinate components with
respect to $c$-independent coordinates, or equivalently its local
connection form with respect to the frame $(v,\e_a)$, have regular
limits (i.e.\ no terms of negative order in the expansion in
$c^{-1}$).  This implies that its local connection form with respect
to $(\E_A)$ expands as
\begin{subequations} \label{eq:Lor_expansion_conn}
\begin{align}
  \tensor{\omegaL}{^0_0} &= 0, \\
  \tensor{\omegaL}{^a_0} &= c^{-1} \varpi^a + \Or(c^{-3})
                           = \delta^{ab} \tensor{\omegaL}{^0_b} \; ,
  \\
  \tensor{\omegaL}{^a_b} &= \tensor{\omega}{^a_b} + \Or(c^{-2})
\end{align}
\end{subequations}
for local one-forms $\tensor{\omega}{^a_b}$, $\varpi^a$.  Under local
rotations and boosts of the frame, the $(\tensor{\omega}{^a_b},
\varpi^a)$ transform as the local connection form of a Galilei
connection on $(M,\tau,h)$ would, thereby in fact defining a Galilei
connection $\bomega$.  Cartan's first structure equation then implies
that the torsion $\TL$ of $\bomegaL$ expands as
\begin{subequations}
\begin{align}
  \TL^0 &= \D\E^0 + \tensor{\omegaL}{^0_a} \wedge \E^a
          = c \D\tau + c^{-1} (\D a + \varpi_a \wedge \e^a)
            + \Or(c^{-3}) \nonumber\\
        &= c \D\tau + c^{-1} f + \Or(c^{-3}), \\
  \TL^a &= \D\E^a + \tensor{\omegaL}{^a_B} \wedge \E^B
          = \D\e^a + \varpi^a \wedge \tau
            + \tensor{\omega}{^a_b} \wedge \e^b + \Or(c^{-2})
          \nonumber\\
        &= T^a + \Or(c^{-2})
\end{align}
\end{subequations}
in terms of the extended torsion of $\bomega$ with respect to the
Bargmann structure $\ba$ obtained from the expansion of the frame.
Assuming $\D\tau = 0$, we can further compute the expansion of the
Lorentzian contortion as
\begin{align}
  \tensor{\KL}{^\rho_{\mu\nu}}
  &= \frac{1}{2} \tensor{\TL}{^\rho_{\mu\nu}}
    - \tensor{\TL}{_{(\mu\nu)}^\rho} \nonumber\\
  &= \frac{1}{2} \underbrace{\tensor{\TL}{^\rho_{\mu\nu}}}
      _{\mathrlap{= \tensor{T}{^\rho_{\mu\nu}} + \Or(c^{-2})}}
    - \eta_{AB} \E^A_{(\mu} \tensor{\TL}{^B_{\nu)\sigma}}
      \underbrace{\eta^{CD} \E_C^\sigma \E_D^\rho}
      _{\mathrlap{= h^{\sigma\rho} + \Or(c^{-2})}} \nonumber\\
  &= \frac{1}{2} \tensor{T}{^\rho_{\mu\nu}}
    + \tau_{(\mu} f_{\nu)\sigma} h^{\sigma\rho}
    - \tensor{h}{_{\kappa(\mu}}
      \tensor{T}{^\kappa_{\nu)\sigma}} h^{\sigma\rho}
    + \Or(c^{-2}) \nonumber\\
  &= \tensor{K}{^\rho_{\mu\nu}} + \Or(c^{-2}).
\end{align}

Finally, we want to comment on the transformation behaviour of the
`limiting' geometric objects on our Galilei manifold under (active)
pushforward of the Lorentzian objects along diffeomorphisms.
Transforming the Lorentzian geometric objects by $c$-independent
diffeomorphisms, all limiting objects also transform simply by
pushforward.  Considering instead `pushforward along an infinitesimal
$c$-dependent diffeomorphism', i.e.\ the transformation
\begin{equation}
  A \to A - c^{-2} \mathcal L_X A + \Or(c^{-4})
\end{equation}
on natural Lorentzian geometric objects $A$ for some vector field $X$,
the only non-trivial transformation of the limiting Galilei-manifold
objects arising from this is the transformation
\begin{equation}
  a \to a - \mathcal L_X \tau = a - \D\tau(X,\cdot) - \D(\tau(X))
\end{equation}
of the local representative of the Bargmann structure.  If the clock
form satisfies our assumption of absolute time, $\D\tau = 0$, this
amounts to a $\Ui$ gauge transformation of the Bargmann structure
\eqref{eq:Barg_Ui_trafo}.  This means that under the assumption of
absolute time, \emph{we obtain all `natural symmetries' of the
  framework of Galilei manifolds with Bargmann structure---%
  diffeomorphisms, local Galilei transformations, and $\Ui$ gauge
  transformations of the Bargmann structure---from the action of
  `$c$-dependent' diffeomorphisms and local Lorentz transformations on
  Lorentzian objects, i.e.\ from the `natural symmetries' of the
  Lorentzian setting}.  For this to hold, the assumption of absolute
time is crucial: otherwise, the limiting geometric objects would
transform not under the Bargmann algebra, but under a certain Lie
algebra expansion of the Poincaré algebra, and would instead define
what has been termed a `torsional Newton--Cartan type \acronym{II}'
(\acronym{TNC} type \acronym{II}) geometry \cite{Hansen.EtAl:2019,
  Hansen.EtAl:2020}.

\subsection{Trace-reversing the field equation of \acronym{TEGR}}

The field equation of \acronym{TEGR} is \cite{Bahamonde.EtAl:2021}
\begin{equation} \label{eq:field_eq_TEGR}
  \E^{-1} \partial_\sigma (\E \tensor{S}{_A^{\mu\sigma}})
  - \tensor{\TL}{^\sigma_{\nu A}} \tensor{S}{_\sigma^{\nu\mu}}
  + \frac{\E^\mu_A}{4} \tensor{S}{_\rho^{\sigma\nu}}
  \tensor{\TL}{^\rho_{\sigma\nu}}
  + \tensor{\omegaL}{_\nu^B_A} \tensor{S}{_B^{\nu\mu}}
  = \frac{8\pi G}{c^4} \tensor{\Theta}{_A^\mu} \; ,
\end{equation}
where $\Theta$ denotes the energy--momentum tensor, $\E =
\det(\E^A_\mu)$ is the determinant of the matrix of dual frame
components, and the superpotential is given by
\begin{equation}
  \tensor{S}{_\rho^{\sigma\mu}}
  = \tensor{\KL}{^\sigma_\rho^\mu}
    - 2 \delta^{[\sigma}_\rho \tensor{\TL}{^{|\nu|}_\nu^{\mu]}} \; .
\end{equation}
Note that we have rewritten the equations such as to conform to our
notation and conventions (in particular regarding the index structure
of the contortion), and inserted the expression of the torsion scalar
in terms of the superpotential \cite[eq.\
(4.160)]{Bahamonde.EtAl:2021} into the field equation \cite[eq.\
(4.163)]{Bahamonde.EtAl:2021}.

In order to consistently take the $c\to\infty$ Newtonian limit of the
field equation, we have to consider it in trace-reversed
form.\footnote{If we did not trace-reverse the equation, we would end
  up with a formal $c^{-1}$ expansion not allowing us to easily
  extract meaningful information about the limit: inserting the
  expansions of the geometric objects as introduced above into the
  field equation, the only order at which the expanded equation would
  make a statement would be identically satisfied; the limit field
  equations proper would appear at the next order, which would no
  longer be contained in the expanded equation since the termination
  of the expansion of the geometric objects introduces unspecified
  terms into the equation.

  I.e.\ to obtain the leading-order equations for the expanded
  geometric objects from the original field equation, one would have
  to expand the objects to higher order than appear in the final
  equations.  This is prevented by considering the trace-reversed
  equation instead.}  The trace of \eqref{eq:field_eq_TEGR}, obtained
by contraction with $E^A_\mu$, is
\begin{equation}
  \E^{-1} \E^A_\mu \partial_\sigma (\E \tensor{S}{_A^{\mu\sigma}})
  + \tensor{\omegaL}{_\nu^B_A} \tensor{S}{_B^{\nu A}}
  = \frac{8\pi G}{c^4} \Theta.
\end{equation}
Rewriting $\E^A_\mu \partial_\sigma (\E \tensor{S}{_A^{\mu\sigma}}) =
\partial_\sigma (\E \tensor{S}{_A^{A\sigma}}) - \E
\tensor{S}{_A^{\mu\sigma}} \partial_\sigma \E^A_\mu$, applying the
identities $\tensor{S}{_A^{A\sigma}} = 2
\tensor{\TL}{^{\nu\sigma}_\nu}$ (which directly follows from the
definition of $S$) and $\partial_\sigma \E^A_\mu =
\GammaL^\kappa_{\sigma\mu} \E^A_\kappa - \tensor{\omegaL}{_\sigma^A_B}
\E^B_\mu$, and using the antisymmetry $\tensor{S}{_A^{\mu\sigma}} =
\tensor{S}{_A^{[\mu\sigma]}}$, the trace equation takes the form
\begin{equation} \label{eq:field_eq_TEGR_trace}
  2 \E^{-1} \partial_\sigma (\E \tensor{\TL}{^{\nu\sigma}_\nu})
  + \frac{1}{2} \tensor{S}{_\rho^{\sigma\nu}}
    \tensor{\TL}{^\rho_{\sigma\nu}}
  = \frac{8\pi G}{c^4} \Theta.
\end{equation}
Now trace-reversing the field equation, i.e.\ considering
$\eqref{eq:field_eq_TEGR} - \frac{1}{2} \E^\mu_A
\eqref{eq:field_eq_TEGR_trace}$, we obtain
\begin{equation} \label{eq:field_eq_TEGR_trace-rev}
  \E^{-1} \partial_\sigma (\E \tensor{S}{_A^{\mu\sigma}})
  - \E^{-1} \E^\mu_A \partial_\sigma
    (\E \tensor{\TL}{^{\nu\sigma}_\nu})
  - \tensor{\TL}{^\sigma_{\nu A}} \tensor{S}{_\sigma^{\nu\mu}}
  + \tensor{\omegaL}{_\nu^B_A} \tensor{S}{_B^{\nu\mu}}
  = \frac{8\pi G}{c^4} (\tensor{\Theta}{_A^\mu}
    - \tfrac{1}{2} \Theta \E^\mu_A).
\end{equation}

We will now further rewrite this equation, in order to express it
purely in terms of the teleparallel connection, the torsion, and the
contortion. Denoting the Lorentzian Levi-Civita connection by
$\nablatildeL$, we have
\begin{equation}
  \E^{-1} \partial_\sigma (\E \tensor{\TL}{^{\nu\sigma}_\nu})
  = \nablatildeL_\sigma \tensor{\TL}{^{\nu\sigma}_\nu}
  = \nablaL_\sigma \tensor{\TL}{^{\nu\sigma}_\nu}
    - \underbrace{\tensor{\KL}{^\sigma_{\sigma\lambda}}}
        _{= \tensor{\TL}{^\sigma_{\sigma\lambda}}}
      \tensor{\TL}{^{\nu\lambda}_\nu} \; ,
\end{equation}
implying
\begin{equation} \label{eq:rewrite_TEGR_field_eq_1}
  \E^{-1} \E^\mu_A \partial_\sigma (\E \tensor{\TL}{^{\nu\sigma}_\nu})
  = \nablaL_\sigma (\E^\mu_A \tensor{\TL}{^{\nu\sigma}_\nu})
    - \tensor{\TL}{^{\nu\sigma}_\nu}
      \underbrace{\nablaL_\sigma \E^\mu_A}
        _{\mathrlap{= \tensor{\omegaL}{_\sigma^B_A} \E^\mu_B}}
    + \E^\mu_A \tensor{\TL}{^\sigma_{\lambda\sigma}}
      \tensor{\TL}{^{\nu\lambda}_\nu} \; ,
\end{equation}
and
\begin{align} \label{eq:rewrite_TEGR_field_eq_2}
  \E^{-1} \partial_\sigma (\E \tensor{S}{_A^{\mu\sigma}})
  &= \underbrace{\frac{\partial_\nu \E}{\E}}
        _{\mathrlap{= \GammatildeL^\sigma_{\sigma\nu}}}
      \tensor{S}{_A^{\mu\nu}}
    + \partial_\sigma \tensor{S}{_A^{\mu\sigma}} \nonumber\\
  &= \nablatildeL_\sigma \tensor{S}{_A^{\mu\sigma}}
    - \GammatildeL^\mu_{\sigma\nu}
      \underbrace{\tensor{S}{_A^{\nu\sigma}}}
        _{= \tensor{S}{_A^{[\nu\sigma]}}} \nonumber\\
  &= \nablatildeL_\sigma \tensor{S}{_A^{\mu\sigma}} \nonumber\\
  &= \nablaL_\sigma \tensor{S}{_A^{\mu\sigma}}
    - \tensor{\KL}{^\mu_{\sigma\nu}} \tensor{S}{_A^{\nu\sigma}}
    - \tensor{\KL}{^\sigma_{\sigma\nu}}
      \tensor{S}{_A^{\mu\nu}} \nonumber\\
  &= \nablaL_\sigma \tensor{S}{_A^{\mu\sigma}}
    - \tfrac{1}{2} \tensor{\TL}{^\mu_{\sigma\nu}}
      \tensor{S}{_A^{\nu\sigma}}
    + \tensor{\TL}{^\sigma_{\nu\sigma}} \tensor{S}{_A^{\mu\nu}} ,
\end{align}
where we used $\tensor{\KL}{^\mu_{[\sigma\nu]}} = \tfrac{1}{2}
\tensor{\TL}{^\mu_{\sigma\nu}}$ and $\tensor{\KL}{^\sigma_{\sigma\nu}}
= - \tensor{\TL}{^\sigma_{\nu\sigma}}$.  Introducing the abbreviation
\begin{equation}
  \tensor{\tilde S}{_A^{\mu\sigma}}
  := \tensor{S}{_A^{\mu\sigma}}
    - \E^\mu_A \tensor{\TL}{^{\nu\sigma}_\nu} \; ,
\end{equation}
we can use \eqref{eq:rewrite_TEGR_field_eq_1} and
\eqref{eq:rewrite_TEGR_field_eq_2} to rewrite the trace-reversed field
equation \eqref{eq:field_eq_TEGR_trace-rev} as
\begin{equation}
  \tensor{\nablaL}{_\sigma} \tensor{\tilde S}{_A^{\mu\sigma}}
  - \tfrac{1}{2} \tensor{\TL}{^\mu_{\sigma\nu}}
    \tensor{S}{_A^{\nu\sigma}}
  + \tensor{\TL}{^\sigma_{\nu\sigma}} \tensor{\tilde S}{_A^{\mu\nu}}
  - \tensor{\TL}{^\sigma_{\nu A}} \tensor{S}{_\sigma^{\nu\mu}}
  - \tensor{\omegaL}{_\sigma^B_A} \tensor{\tilde S}{_B^{\mu\sigma}}
  = \frac{8\pi G}{c^4} (\tensor{\Theta}{_A^\mu}
    - \tfrac{1}{2} \Theta \E^\mu_A).
\end{equation}
Using $\tensor{\nablaL}{_\sigma} \tensor{\tilde S}{_A^{\mu\sigma}} -
\tensor{\omegaL}{_\sigma^B_A} \tensor{\tilde S}{_B^{\mu\sigma}} =
\tensor{\DL}{_\sigma} \tensor{\tilde S}{_A^{\mu\sigma}}$, lowering the
index $\mu$, inserting the identity $\tensor{\tilde S}{_{A\mu}^\sigma}
= - \tensor{\KL}{^\sigma_{\mu A}} - \E^\sigma_A
\tensor{\TL}{^\nu_{\mu\nu}} = - \tensor{\KL}{^\sigma_{\mu A}} +
\E^\sigma_A \tensor{\KL}{^\nu_{\nu\mu}}$ as well as the definition of
$S$, and contracting with $\E^\mu_B$, this equation can be shown to be
equivalent to
\begin{equation} \label{eq:field_eq_TEGR_trace-rev_final}
  - \tensor{\DL}{_\sigma} \tensor{\KL}{^\sigma_{AB}}
  + \tensor{\DL}{_A} \tensor{\KL}{^\mu_{\mu B}}
  - \tensor{\KL}{^\mu_{\sigma B}} \tensor{\TL}{^\sigma_{\mu A}}
  + \tensor{\KL}{^\mu_{\mu\sigma}} \tensor{\KL}{^\sigma_{AB}}
  - \tensor{\KL}{^\mu_{A\sigma}} \tensor{\KL}{^\sigma_{\mu B}}
  = \frac{8\pi G}{c^4} (\Theta_{AB} - \tfrac{1}{2} \Theta \, \eta_{AB}).
\end{equation}
Note that we could have arrived at this result more directly, or at
least anticipated it beforehand: the left-hand side of
\eqref{eq:field_eq_TEGR_trace-rev_final} is simply the component
$\accentset{\text{L}}{\widetilde R}_{BA}$ of the Ricci tensor of the
Lorentzian Levi-Civita connection expressed in terms of the
teleparallel connection, its torsion, and the contortion, such that
the equation is simply the trace-reversed Einstein equation.  However,
we wanted to keep the argumentation `as teleparallel as possible' and
therefore have presented the calculation leading to
\eqref{eq:field_eq_TEGR_trace-rev_final} with as little reference to
the Levi-Civita connection as possible.

\subsection{Taking the limit}

We are now going to expand the field equation
\eqref{eq:field_eq_TEGR_trace-rev_final} in $c^{-1}$, and thus have to
expand the trace-reversed energy--momentum tensor appearing on its
right-hand side.  First working in the Lorentzian setting without any
$c$-expansion, we may decompose the energy--momentum tensor with
respect to any unit future-directed timelike vector field $\xi$ as
\begin{equation}
  \Theta_{\mu\nu} = \xi_\mu \xi_\nu w - c 2 \xi_{(\mu} \Pi_{\nu)}
    + \Sigma_{\mu\nu} \; ,
\end{equation}
where $w$, $\Pi$ and $\Sigma$ are the energy density, the momentum
density and the momentum flux density (stress tensor) with respect to
$\xi$.  Choosing $\xi^\mu = \E_0^\mu$ from our tetrad, which implies
$\xi_\mu = - \E^0_\mu$, and expanding the energy density in this frame
as
\begin{equation}
  w = \rho c^2 + w^{(0)} + \Or(c^{-2})
\end{equation}
with $\rho$ being the (rest) mass density, while taking momentum
density and momentum flux density to be of order $c^0$, we obtain the
expansion
\begin{equation}
  \Theta_{\mu\nu} = c^4 \rho \tau_\mu \tau_\nu
    + c^2 w^{(0)} \tau_\mu \tau_\nu + c^2 2 \rho \tau_{(\mu} a_{\nu)}
    + c^2 2 \tau_{(\mu} \Pi_{\nu)} + \Or(c^0)
\end{equation}
for the energy--momentum tensor, where we have used the expansion
\eqref{eq:Lor_expansion_tetr} of the tetrad.\footnote{Note that the
  mass density $\rho$ is invariant under changes of frame.}
Contracting this expansion again with that of the tetrad components
\eqref{eq:Lor_expansion_tetr}, we obtain the trace $\Theta = - c^2
\rho + \Or(c^0)$.  Thus, the trace-reversed energy--momentum tensor is
\begin{align}
  \Theta_{\mu\nu} - \tfrac{1}{2} \Theta \E^A_\mu \E^B_\nu \eta_{AB}
  &= \Theta_{\mu\nu} - \tfrac{1}{2} \Theta
    (- c^2 \tau_\mu \tau_\nu + \Or(c^0)) \nonumber\\
  &= \tfrac{1}{2} c^4 \rho \tau_\mu \tau_\nu + \Or(c^2),
\end{align}
and the expansion of the right-hand side of the trace-reversed
\acronym{TEGR} field equation \eqref{eq:field_eq_TEGR_trace-rev_final}
is $4\pi G \, \rho \, \tau_A \tau_B + \Or(c^{-2})$.  Thus, expanding
the objects on the left-hand side as well, we see that in the
$c\to\infty$ limit, the \acronym{TEGR} field equation goes over to the
field equation \eqref{eq:field_eq_telNC} of teleparallel
Newton--Cartan gravity.

Regarding curvature, one easily checks that the components of the
curvature form of any Lorentzian connection expanded as in
\eqref{eq:Lor_expansion_conn} take the form
\begin{subequations}
\begin{align}
  \tensor{\RL}{^a_0} &= c^{-1} (\D\varpi^a
        + \omega^a_b \wedge \varpi_b) + \Or(c^{-3})
    = c^{-1} \tensor{R}{^a_t} + \Or(c^{-3}), \\
  \tensor{\RL}{^a_b} &= \D\omega^a_b
      + \omega^a_c \wedge \omega^c_b + \Or(c^{-2})
    = \tensor{R}{^a_b} + \Or(c^{-2}), \\
  \tensor{\RL}{^0_0} &= \Or(c^{-4}), \\
  \tensor{\RL}{^0_a} &= c^{-1} \delta_{ab}
      \tensor{R}{^b_t} + \Or(c^{-3})
\end{align}
\end{subequations}
in terms of the curvature form of the limiting Galilei connection.
Therefore, if $\omegaL$ is flat, then also the limiting Galilei
connection $\omega$ will be flat.

Finally, it is clear that the \acronym{TEGR} test particle equation of
motion
\begin{equation}
  (\nabla_{\dot\gamma} \dot\gamma)^\rho
  = \tensor{\KL}{^\rho_{\mu\nu}} \dot\gamma^\mu \dot\gamma^\nu
\end{equation}
goes over to its teleparallel Newton--Cartan equivalent
\eqref{eq:EoM_telNC} in the $c\to\infty$ limit.  Thus, we have shown
that the formal $c\to\infty$ limit of \acronym{TEGR} with respect to a
closed clock one-form $\tau$ is teleparallel Newton--Cartan gravity as
introduced in section \ref{sec:Barg_structs_telep_Gal_conns}.

\section{Recovering Newtonian gravity}
\label{sec:recover_Newton}

As for usual Newton--Cartan gravity, the standard formulation of
Newtonian gravity can be recovered from teleparallel Newton--Cartan
gravity, as we will show now.

\subsection{The field equation}

The $tt$-component of the teleparallel Newton--Cartan field equation
\eqref{eq:field_eq_telNC}, i.e.\ the equation for $A = B = t$, can be
rewritten as
\begin{equation} \label{eq:field_eq_tt}
  - \tensor{D}{_\sigma} \tensor{K}{^\sigma_{tt}}
  - \tensor{D}{_t} \tensor{T}{^\mu_{t\mu}}
  - \tensor{K}{^a_{bt}} \tensor{T}{^b_{at}}
  + \tensor{T}{^a_{ab}} \tensor{K}{^b_{tt}}
  - \tensor{K}{^a_{tb}} \tensor{K}{^b_{at}}
  = 4\pi G \rho ,
\end{equation}
where we used that $T^t = 0$ and therefore $\tensor{K}{^t_{\mu\nu}} =
0$.  In the following, we are going to show how to recover from this
the standard formulation of the Newtonian field equation.

As a first step, we `gauge-fix' the connection $\bomega$ to vanishing
purely spatial torsion, i.e.\ we assume
\begin{equation} \label{eq:spat_torsion_vanish}
  \tensor{T}{^a_{bc}} = 0.
\end{equation}
Here we put the term `gauge-fix' in quotation marks since we do
\emph{not} use it to refer to the fixing of any gauge redundancy in
the proper sense (e.g.\ in a Hamiltonian analysis).  Instead, we only
use it to mean that we add \eqref{eq:spat_torsion_vanish} as an
additional assumption on the connection $\bomega$, which is compatible
with the field equations and does not introduce any further
restrictions.  That the choice of vanishing purely spatial torsion is
always possible follows from the purely spatial components ($A = a$,
$B = b$) of the field equation \eqref{eq:field_eq_telNC}: this part of
the equation reads $\widetilde R_{ab} = 0$ in terms of the Ricci
tensor of the extended-torsion-free connection
$\boldsymbol{\widetilde\omega}$.  This means that the spatial leaves
are Ricci-flat as Riemannian manifolds (with the spatial metric
induced by $h$); since they are three-dimensional, this implies that
they are flat.  Therefore, by choosing the spatial connection
$\tensor{\omega}{^a_b}$ to be the Levi-Civita connection
$\tensor{\widetilde\omega}{^a_b}$ of the spatial leaves, we may indeed
satisfy \eqref{eq:spat_torsion_vanish} as well as flatness of
$\bomega$.

Let us stress here again that this `gauge-fixing' assumption of
vanishing purely spatial torsion is, differently to the situation
considered in \cite{Read.Teh:2018}, \emph{not} part of the formulation
of the theory, but only added \emph{afterwards} for the recovery of
standard Newtonian gravity.  Note also that it is more general than
the torsion constraint from \cite{Read.Teh:2018}, where the
\emph{total} spatial torsion $\tensor{T}{^a_{\mu\nu}}$ is assumed to
vanish (i.e.\ including its mixed spatio-temporal components
$\tensor{T}{^a_{tb}}$).\footnote{In \cite{Read.Teh:2018}, this torsion
  constraint is not even emphasised as an assumption, but instead it
  is stated that `for a flat connection, the Bianchi identities imply
  that [the spatial torsion vanishes]'.  This clearly is not true at
  all; one can easily give examples of flat metric torsionful
  connections in, for example, the 2-dimensional Euclidean plane.}

Now we are going to rewrite \eqref{eq:field_eq_tt}.  From
\begin{equation}
  \tensor{K}{^\sigma_{(AB)}}
  = \tensor{\Ksp}{^\sigma_{(AB)}}
    + \tensor{\tau}{_{(A}} \tensor{f}{_{B)}^\sigma}
  = - \tensor{T}{_{(AB)}^\sigma}
    + \tensor{\tau}{_{(A}} \tensor{f}{_{B)}^\sigma}
\end{equation}
we obtain
\begin{align} \label{eq:rewrite_field_eq_1}
  \tensor{D}{_\sigma} \tensor{K}{^\sigma_{tt}}
  &= - \tensor{D}{_\sigma} \tensor{T}{_{tt}^\sigma}
    + \tensor{D}{_\sigma} \tensor{f}{_t^\sigma} \nonumber\\
  &= \tensor{\varpi}{_\sigma^a} \tensor{T}{_{at}^\sigma}
    + \tensor{D}{_a} \tensor{f}{_t^a} \nonumber\\
  &= - \tensor{\varpi}{_b^a} \tensor{T}{_a^b_t}
    - \tensor{D}{_a} \tensor{f}{^a_t} \; ,
\end{align}
where at the second equality sign we used $T_{t\mu\nu} = 0$ for the
first and \eqref{eq:cov_deriv_t_zero} for the second term.
Furthermore, from
\begin{equation}
  \tensor{D}{_A} \tensor{T}{^\mu_{B\mu}}
  = \tensor{\partial}{_A} \tensor{T}{^\mu_{B\mu}}
  - \tensor{\omega}{_A^C_B} \tensor{T}{^\mu_{C\mu}}
\end{equation}
we obtain
\begin{equation} \label{eq:rewrite_field_eq_2}
  \tensor{D}{_t} \tensor{T}{^\mu_{t\mu}}
  = \tensor{\partial}{_t} \tensor{T}{^\mu_{t\mu}}
    - \tensor{\varpi}{_t^c} \tensor{T}{^\mu_{c\mu}}
  = - \tensor{\partial}{_t} \tensor{T}{^a_{at}} \; ,
\end{equation}
where we used our `gauge assumption' $\tensor{T}{^a_{bc}} = 0$.  A
direct calculation further shows that
\begin{equation} \label{eq:rewrite_field_eq_3}
  - \tensor{K}{^a_{bt}} \tensor{T}{^b_{at}}
    - \tensor{K}{^a_{tb}} \tensor{K}{^b_{at}}
  = - \tensor{T}{^{ab}_t} \tensor{T}{_{(ab)t}}
    + \frac{1}{4} \tensor{f}{^{ab}} \tensor{f}{_{ab}} \; .
\end{equation}
Using \eqref{eq:rewrite_field_eq_1}, \eqref{eq:rewrite_field_eq_2},
\eqref{eq:rewrite_field_eq_3} and (again) vanishing of the purely
spatial torsion, we can rewrite the $tt$-component
\eqref{eq:field_eq_tt} of the field equation as
\begin{equation} \label{eq:field_eq_tt_no_purely_spatial_torsion}
  \tensor{D}{_a} \tensor{f}{^a_t}
  + \tensor{\varpi}{_b^a} \tensor{T}{_a^b_t}
  + \tensor{\partial}{_t} \tensor{T}{^a_{at}}
  - \tensor{T}{^{ab}_t} \tensor{T}{_{(ab)t}}
  + \frac{1}{4} \tensor{f}{^{ab}} \tensor{f}{_{ab}}
  = 4\pi G \rho.
\end{equation}

From now on, we are going to assume the `absolute rotation' condition
introduced by Trautman \cite{Trautman:1963,Trautman:1965}.  This is an
additional curvature condition that has to be assumed in usual
Newton--Cartan gravity in order to recover Newtonian gravity in the
usual sense, taking the form
\begin{equation} \label{eq:abs_rot}
  \tensor{\widetilde R}{^{\mu\nu}_{\rho\sigma}} = 0.
\end{equation}
Without assuming this, Newton--Cartan gravity actually describes a
slightly generalised version of Newtonian gravity, in which the
vorticity of rigid timelike flows need not be spatially constant.  If
in a Newton--Cartan spacetime there is one rigid timelike vector field
with spatially variable vorticity, then actually \emph{all} rigid
timelike vector fields have spatially non-constant vorticity, and in
this sense in such a spacetime no absolute notion of rotation exists.
In the `recovered' Newtonian equations this leads to non-eliminable
Coriolis force terms.  If one however \emph{assumes} the absolute
rotation condition \eqref{eq:abs_rot}, rigid non-rotating frames
exist, and in those one recovers Newtonian gravity
proper.\footnote{The interpretation of the additional curvature
  condition \eqref{eq:abs_rot} in terms of rotation was, as far as I
  (the author) know, not given by Trautman, at least not explicitly.
  To my knowledge, it first appears in explicit form in Ehlers' 1981
  article on frame theory \cite{Ehlers:1981a,Ehlers:1981b}.}  Note
that \eqref{eq:abs_rot} also implies $\tensor{\widetilde R}{^a_t} =
0$, i.e.\ the $at$-component of the field equation
\eqref{eq:field_eq_telNC}.

Assuming absolute rotation, we now choose the timelike frame field $v
= \e_t$ to be rigid and non-rotating.  In usual Newton--Cartan
gravity, i.e.\ in terms of the extended-torsion-free connection, this
means that the frame field satisfies $\mathcal L_v h = 0$ or
equivalently $\widetilde\nabla^{(\mu} v^{\nu)} = 0$ (rigidness), as
well as $\widetilde\nabla^{[\mu} v^{\nu]} = 0$ (it is non-rotating).
Expressed in terms of our `teleparallel' Newton--Cartan connection
$\nabla$, this means\footnote{This may be verified by direct
  computation, using the rigidness and non-rotation conditions in
  terms of $\widetilde\nabla$ and the explicit form of the
  Newton--Cartan contortion \eqref{eq:defn_NC_contort}.}
\begin{subequations} \label{eq:Trautman_frame}
\begin{align}
  \tensor{\varpi}{_{(ab)}} &= \tensor{T}{_{(ab)t}}
    \; , \label{eq:Trautman_frame_sym} \\
  \tensor{\varpi}{_{[ab]}} &= \frac{1}{2} \tensor{f}{_{ab}}
    \; . \label{eq:Trautman_frame_antisym}
\end{align}
\end{subequations}
The second of these equations is equivalent to $(\D a)_{ab} = 0$,
i.e.\ $0 = (\D a)|_{\Sigma} = \D(a|_{\Sigma})$ on any spatial leaf
$\Sigma$ (where $a$ is the frame representative of the Bargmann
structure).  Thus, we locally have $a|_{\Sigma} = \D u$ for some
function $u$ on $\Sigma$; choosing $u$ smoothly between leaves, we
thus have $a = \D u + \phi \tau$ for functions $u,\phi$.  In
particular, we have
\begin{equation} \label{eq:def_phi_Trautman_frame}
  \D a = \D\phi \wedge \tau.
\end{equation}
Obviously, this conversely implies $(\D a)_{ab} = 0$, i.e.\
\eqref{eq:Trautman_frame_antisym}.  The (locally defined,
frame-dependent) function $\phi$ defined by
\eqref{eq:def_phi_Trautman_frame} will play the rôle of the Newtonian
potential, and is defined up to addition of a time-dependent spatially
constant function.\footnote{Note that this is precisely the same as
  for usual Newton--Cartan gravity: there, the Newtonian potential in
  the reconstruction of Newtonian gravity in a rigid non-rotating
  frame is defined by $\tilde\Omega = - \D\phi \wedge \tau$.}

From now on working in a rigid non-rotating frame in the sense
introduced above, we have $f = \D a + \Omega = \D\phi \wedge \tau +
\Omega$, in particular implying
\begin{equation} \label{eq:f_at_Trautman_frame}
  \tensor{f}{_{at}}
  = \tensor{\partial}{_a} \phi
    + 2 \tensor{\varpi}{_{[at]}}
  = \tensor{\partial}{_a} \phi
    + \cancel{\tensor{\varpi}{_{at}}}
    - \tensor{\varpi}{_{ta}} \; .
\end{equation}
Thus, we can rewrite
\begin{align} \label{eq:rewrite_field_eq_Trautman_frame_1}
  \tensor{D}{_a} \tensor{f}{^a_t}
  &= \tensor{\partial}{_a} \tensor{f}{^a_t}
    + \tensor{\omega}{_a^a_b} \tensor{f}{^b_t}
    - \underbrace{\tensor{\varpi}{_a^b} \tensor{f}{^a_b}}
      _{\mathrlap{\overset{\eqref{eq:Trautman_frame_antisym}}{=}
        2 \tensor{\varpi}{^{[ab]}} \tensor{\varpi}{_{[ab]}}}}
    \nonumber\\
  &= \underbrace{\tensor{\partial}{_a} \tensor{\partial}{^a} \phi
    + \tensor{\omega}{_a^a_b} \tensor{\partial}{^b} \phi}
      _{= \tensor{D}{_a} \tensor{D}{^a} \phi}
    - \tensor{\partial}{_a} \tensor{\varpi}{_t^a}
    - \tensor{\omega}{_a^a_b} \tensor{\varpi}{_t^b}
    - 2 \tensor{\varpi}{^{[ab]}} \tensor{\varpi}{_{[ab]}} \; .
\end{align}

For any one-form $\alpha$, we have the general identity
\begin{equation}
  \tensor{(\D\alpha)}{_{\mu\nu}}
  = 2 \tensor{\partial}{_{[\mu}} \tensor{\alpha}{_{\nu]}}
  = 2 \tensor{\nabla}{_{[\mu}} \tensor{\alpha}{_{\nu]}}
    + \tensor{T}{^\rho_{\mu\nu}} \tensor{\alpha}{_\rho} \; ,
\end{equation}
giving in particular
\begin{align}
  \tensor{(\D\alpha)}{_{at}}
  &= 2 \tensor{D}{_{[a}} \tensor{\alpha}{_{t]}}
    + \tensor{T}{^B_{at}} \tensor{\alpha}{_B} \nonumber\\
  &= 2 \tensor{\partial}{_{[a}} \tensor{\alpha}{_{t]}}
    - \tensor{\varpi}{_a^b} \tensor{\alpha}{_b}
    + \tensor{\omega}{_t^b_a} \tensor{\alpha}{_b}
    + \tensor{T}{^b_{at}} \tensor{\alpha}{_b} \; .
\end{align}
Applied to the one-form $\varpi^a$, we obtain
\begin{equation}
  \tensor{(\D\varpi^a)}{_{at}}
  = 2 \tensor{\partial}{_{[a}} \tensor{\varpi}{_{t]}^a}
    - \underbrace{\tensor{\varpi}{_a^b} \tensor{\varpi}{_b^a}}
      _{= \tensor{\varpi}{^{ab}} \tensor{\varpi}{_{ba}}}
    + \tensor{\omega}{_t^b_a} \tensor{\varpi}{_b^a}
    + \tensor{T}{^b_{at}} \tensor{\varpi}{_b^a} \; .
\end{equation}
Flatness of the connection implies $0 = \D\tensor{\varpi}{^a} +
\tensor{\omega}{^a_b} \wedge \tensor{\varpi}{^b}$, which combined with
the previous equation leads to
\begin{align} \label{eq:rewrite_field_eq_Trautman_frame_2}
  0 &= (\D\tensor{\varpi}{^a})_{at} + (\tensor{\omega}{^a_b}
      \wedge \tensor{\varpi}{^b})_{at} \nonumber\\
    &= \tensor{\partial}{_a} \tensor{\varpi}{_t^a}
      - \tensor{\partial}{_t} \underbrace{\tensor{\varpi}{_a^a}}
        _{\mathclap{\overset{\eqref{eq:Trautman_frame_sym}}{=}
          \tensor{T}{^a_{at}}}}
      \!{}- \tensor{\varpi}{^{ab}} \tensor{\varpi}{_{ba}}
      + \cancel{\tensor{\omega}{_t^b_a} \tensor{\varpi}{_b^a}}
      + \tensor{T}{^b_{at}} \tensor{\varpi}{_b^a}
      + \tensor{\omega}{_a^a_b} \tensor{\varpi}{_t^b}
      - \cancel{\tensor{\omega}{_t^a_b} \tensor{\varpi}{_a^b}}.
\end{align}

Combining \eqref{eq:rewrite_field_eq_Trautman_frame_1} and
\eqref{eq:rewrite_field_eq_Trautman_frame_2}, several terms cancel out
and we arrive at
\begin{equation}
  \tensor{D}{_a} \tensor{f}{^a_t}
    + \tensor{\partial}{_t} \tensor{T}{^a_{at}}
  = \tensor{D}{_a} \tensor{D}{^a} \phi
    \underbrace{\!{}- 2 \tensor{\varpi}{^{[ab]}}
      \tensor{\varpi}{_{[ab]}}
    - \tensor{\varpi}{^{ab}} \tensor{\varpi}{_{ba}}}
      _{= - \tensor{\varpi}{^{ab}} \tensor{\varpi}{_{ab}}}
    + \tensor{T}{^b_{at}} \tensor{\varpi}{_b^a}.
\end{equation}
Inserting this, we can rewrite the $tt$-component
\eqref{eq:field_eq_tt_no_purely_spatial_torsion} of the field equation
as
\begin{equation}
  \tensor{D}{_a} \tensor{D}{^a} \phi
  - \tensor{\varpi}{^{ab}} \tensor{\varpi}{_{ab}}
  + \tensor{\varpi}{_{(ba)}} 2 \tensor{T}{^{(ab)}_t}
  - \tensor{T}{^{ab}_t} \tensor{T}{_{(ab)t}}
  + \frac{1}{4} \tensor{f}{^{ab}} \tensor{f}{_{ab}}
  = 4\pi G \rho,
\end{equation}
which using the conditions \eqref{eq:Trautman_frame} on the frame
reduces to
\begin{equation}
  \tensor{D}{_a} \tensor{D}{^a} \phi = 4\pi G \rho,
\end{equation}
the Newtonian field equation.  Note that due to our choice of
vanishing purely spatial torsion, the induced spatial connection is
the Levi-Civita connection of the induced spatial metric, such that
the expression $\tensor{D}{_a} \tensor{D}{^a} \phi$ is the spatial
metric Laplace operator acting on $\phi$.

\subsection{The test particle equation of motion}

Of course, to warrant the claim that we can recover Newtonian gravity,
we not only have to show that we have a field $\phi$ satisfying the
Newtonian field equation, but also that test particles couple to it in
the correct way.  We will now show how the test particle equation of
motion may be reduced to its usual Newtonian counterpart.

We consider the equation of motion \eqref{eq:EoM_telNC} for a unit
timelike vector field $\xi$, i.e.
\begin{equation}
  \xi^\sigma \nabla_\sigma \xi^\rho
  = \tensor{K}{^\rho_{\mu\nu}} \xi^\mu \xi^\nu ,
  \quad \tau_\mu \xi^\mu = 1.
\end{equation}
Inserting the definition of $K$, this equation is equivalent to
\begin{equation}
  \xi^\sigma D_\sigma \xi^a
  = (- \tensor{T}{_{bt}^a} + \tensor{f}{_b^a}) \xi^b
    + \tensor{f}{_t^a}.
\end{equation}
Now in a rigid, non-rotating frame \eqref{eq:Trautman_frame} with
$\phi$ defined by \eqref{eq:def_phi_Trautman_frame}, we have
\begin{equation}
  - \tensor{T}{_{bt}^a} = \tensor{T}{_b^a_t}
  \overset{\eqref{eq:Trautman_frame_sym}}{=}
  \tensor{\varpi}{^a_b} + \tensor{\varpi}{_b^a} - \tensor{T}{^a_{bt}}
  \; .
\end{equation}
Using this, \eqref{eq:Trautman_frame_antisym} and
\eqref{eq:f_at_Trautman_frame}, we can rewrite the equation of motion
as follows:
\begin{align} \label{eq:EoM_rewritten_1}
  \xi^\sigma D_\sigma \xi^a
  &= 2 \xi^b \tensor{\varpi}{_b^a} + \tensor{\varpi}{_t^a}
    - \tensor{T}{^a_{bt}} \xi^b - \partial^a \phi \nonumber\\
  &= \xi^\sigma \tensor{\varpi}{_\sigma^a} + (\tensor{\varpi}{_b^a}
    - \tensor{T}{^a_{bt}}) \xi^b - \partial^a \phi
\end{align}

We now specialise to a frame such that the spacelike frame fields
commute with the timelike one, i.e.\ $[\e_a, v] = 0$.  Such a frame
may always be constructed by starting with orthonormal vector fields
$\e_a$ on one spatial leaf and then extending those along the flow of
$v$ to vector fields on spacetime.  The above commutation requirement
is equivalent to $T(\e_b, v) = \nabla_{\e_b} v - \nabla_v \e_b$, or in
components to
\begin{equation}
  \tensor{\varpi}{_b^a} - \tensor{T}{^a_{bt}}
  = \tensor{\omega}{_t^a_b}
\end{equation}
(where we used $T^t = 0$).  Thus, in such a frame the equation of
motion \eqref{eq:EoM_rewritten_1} takes the form
\begin{equation}
  \xi^\sigma D_\sigma \xi^a
  = \xi^\sigma \tensor{\varpi}{_\sigma^a}
    + \tensor{\omega}{_t^a_b} \xi^b
    - \partial^a \phi.
\end{equation}
Explicitly expressing the left-hand side in terms of components, we
obtain
\begin{equation}
  \xi^\sigma \partial_\sigma \xi^a
    + \tensor{\omega}{_c^a_b} \xi^c\xi^b
  = - \partial^a \phi,
\end{equation}
which for a flow line $\gamma$ of $\xi$ becomes\footnote{Note that due
  to $\tau_\mu \xi^\mu = 1$, the parameter of $\gamma$ is Newtonian
  time $t$ as defined by $\tau$.}
\begin{equation}
  \ddot\gamma^a(t)
    + \tensor{\omega}{_c^a_b}(\gamma(t)) \:
      \dot\gamma^c(t) \: \dot\gamma^b(t)
  = - \partial^a \phi (\gamma(t)).
\end{equation}
This is the Newtonian equation of motion for a test particle,
expressed in a non-Cartesian orthonormal spatial frame.

\section{Conclusion}
\label{sec:conclusion}

In this paper, we have shown how the local formulation of the geometry
of Galilei spacetimes in terms of the Bargmann group gives rise to a
natural notion of teleparallel Galilei connections, allowing for a
teleparallel formulation of Newton--Cartan gravity that generalises
the special case from \cite{Read.Teh:2018}.  We have shown that this
theory is the natural $c \to \infty$ limit of \acronym{TEGR} (with
respect to a closed clock form / leading order timelike dual frame
field), and explained how to recover standard Newtonian gravity from
it.  Thus, teleparallel Newton--Cartan gravity provides a geometric
description of how \acronym{TEGR} gives rise to Newtonian gravity as
its Newtonian limit, in the same way as usual Newton--Cartan gravity
does this for general relativity.

The work presented here lends itself to several interesting
generalisations.  First, one may wonder about a generalisation to the
situation of so-called \emph{torsional Newton--Cartan gravity}
(\acronym{TNC} gravity) \cite{Christensen.EtAl:2014a,
  Christensen.EtAl:2014b}, i.e.\ to a non-closed clock form $\D\tau
\ne 0$, allowing for non-absolute time.  Note, however, that in order
to obtain such a theory as a (formal) limit of \acronym{TEGR} (with
respect to a non-closed $\tau$), one probably would have to take as
the `gauged' algebra that locally describes the geometry not the
Bargmann algebra, but a specific Lie algebra expansion of the Poincaré
algebra.  This means that one would have to consider what has been
termed a `torsional Newton--Cartan type \acronym{II}' (\acronym{TNC}
type \acronym{II}) geometry instead of a \acronym{TNC} type
\acronym{I} geometry \cite{Hansen.EtAl:2019,Hansen.EtAl:2020}.

A second direction for further investigations concerns the formulation
of teleparallel Newton--Cartan gravity in terms of an action
principle, instead of the sole consideration of the equations of
motion as done here.  Since in particular \emph{modified} teleparallel
theories of gravity are commonly formulated in terms of an action
principle, this would enable the investigation of the geometric
Newtonian limit of modified teleparallel theories of gravity.  Note
that similar to the previous point, for developing an action principle
one probably needs to consider \acronym{TNC} type \acronym{II}
geometry, since standard Newton--Cartan gravity and \acronym{TNC} type
\acronym{I} gravity cannot be given a variational formulation
\cite{Hansen.EtAl:2019}.

One may also seek modifications of the present work describing the
Newtonian limit of theories of gravity based on further different
geometries.  For example, one could consider so-called \emph{symmetric
  teleparallel} gravity theories, in which the `Lorentzian' connection
is flat and torsion-free but has non-metricity, or analyse even more
general metric-affine theories in which the connection can have some
combination of curvature, torsion and non-metricity.

Finally, the formulation of teleparallel Newton--Cartan gravity allows
for the investigation of the post-Newtonian expansion of (modified)
teleparallel gravity theories in a geometric, coordinate-free way,
complementing the coordinate approach from \cite{Emtsova.Hohmann:2019,
  Ualikhanova.Hohmann:2019,Hohmann:2021}.  Such a geometric
description of the post-Newtonian expansion of standard general
relativity, starting with and going beyond usual Newton--Cartan
gravity, has been developed in
\cite{Dautcourt:1997,Tichy.Flanagan:2011} and widely extended in
\cite{Hansen.EtAl:2019,Hansen.EtAl:2020} (although the latter
references take a somewhat more `field-theoretic' / gauge-theoretic
perspective on the topic).  The development of such a geometric
description of the post-Newtonian expansion of modified (teleparallel)
theories of gravity we view as the most interesting direction for
further research, since, to quote the introduction, questions
regarding an inherently geometric theory ought to be answered in a
geometric fashion.\footnote{Of course, for practical computations in
  post-Newtonian theory the standard coordinate approach has some
  clear advantages.  Nevertheless, from a conceptual point of view a
  geometric understanding is of fundamental importance, and it may
  also offer insights for concrete calculations, e.g.\ regarding
  questions of coordinate (in-)dependence of observed phenomena.}

\section*{Acknowledgements}

I wish to thank Domenico Giulini for valuable discussions, and Lennart
Janshen, Christian Pfeifer and Sascha Gehrmann for providing helpful
comments on the manuscript.

\nocite{apsrev41Control}
\bibliography{telepar_NC,revtex-custom}

\appendix

\section{Semidirect extension of principal bundles}
\label{sec:appendix_semidir_extension}

Here we will explain the `semidirect extension' construction for
principal bundles and connections on them that is used in the main
text in the definition of Bargmann structures in section
\ref{subsec:Barg_structs}.

Let $H, N$ be Lie groups and $\rho \colon H \to \Aut(N)$ a smooth
homomorphism.  By $\dot\rho \colon H \to \Aut (\mathfrak{n})$, we
denote the induced representation of $H$ on $\mathfrak{n}$ by Lie
algebra automorphisms, defined by
\begin{equation}
  \dot\rho_h := \DD(\rho_h)|_{\e_N} \in \mathsf{Aut}(\mathfrak{n}).
\end{equation}
Let $P \overset{\pi}{\to} M$ be a principal $H$-bundle.  We can extend
$P$ to a principal $H\ltimes N$-bundle $Q \overset{\hat\pi}{\to} M$ as
follows: denoting by $\tilde\rho \colon H \to \Diff(H\ltimes N)$ the
natural left action of $H$ on $H\ltimes N$ by multiplication, i.e.
\begin{subequations}
\begin{equation}
  \tilde\rho_{h_2}(h_1,n) := (h_2, \e_N) (h_1, n)
  = (h_2 h_1, \rho_{h_2}(n)),
\end{equation}
we define $Q$ as the associated bundle
\begin{equation}
  Q := P \times_{\tilde\rho} (H\ltimes N).
\end{equation}
\end{subequations}
The natural right action of $H\ltimes N$ on itself induces a free
right action on $Q$ which is transitive on the fibres and compatible
with the local trivialisations, thus making $Q$ into a principal
bundle as desired.  We also obtain natural bundle homomorphisms $Q
\overunderset{\beta} {\gamma}{\rightleftarrows} P$ satisfying $\beta
\circ \gamma = \mathrm{id}_P$, namely
\begin{equation}
  \gamma(p) = [p, (\e_H, \e_N)], \quad \beta([p, (h,n)]) = ph \; .
\end{equation}
By construction, with respect to local trivialisations $P
\overset{\gamma}{\to} Q$ looks like the inclusion $H \hookrightarrow H
\ltimes N$, such that it really exhibits $Q$ as an extension of $P$.

\begin{theorem}
  Let $\bhatomega \in \Omega^1(Q, \mathfrak{h} \oright \mathfrak{n})$
  be a connection on $Q$.  We decompose its pullback along $\gamma$ as
  $\gamma^*\bhatomega = (\bomega, \btheta)$ with $\bomega \in
  \Omega^1(P, \mathfrak{h})$ and $\btheta \in \Omega^1(P,
  \mathfrak{n})$.  Then $\bomega$ is a connection and $\btheta$ is a
  $\dot\rho$-tensorial form on $P$.

  Conversely, given a connection $\bomega \in \Omega^1(P,
  \mathfrak{h})$ and a $\btheta \in \Omega^1_{\dot\rho}(P,
  \mathfrak{n})$, there is a unique connection $\bhatomega \in
  \Omega^1(Q, \mathfrak{h} \oright \mathfrak{n})$ such that
  $\gamma^*\bhatomega = (\bomega, \btheta)$.

  The curvature form $\boldsymbol{\hat R} \in \Omega^2 (Q,
  \mathfrak{h} \oright \mathfrak{n})$ of $\bhatomega$ satisfies
  \begin{equation}
    \gamma^*\boldsymbol{\hat R}
    = (\boldsymbol{R}, \D^\bomega \btheta
      + \tfrac{1}{2} [\btheta \wedge \btheta]) ,
  \end{equation}
  where $\boldsymbol{R} \in \Omega^2(p, \mathfrak{h})$ is the
  curvature form of $\bomega$.

  \begin{proof}[Idea of proof]
    Using the explicit form of the adjoint representation of a
    semidirect product group, the $\mathrm{Ad}$-equivariance of
    $\bhatomega$, and the $H$-equivariance of $\gamma$, one can check
    the properties needed of $\bomega$ and $\btheta$ by direct
    computation.

    For the converse, $\bhatomega$ is determined on the image of
    $\gamma$ in the `$N$ directions' of $TQ$, which are generated by
    fundamental vector fields, by being a connection, and in the
    remaining directions by its pullback $\gamma^*\bhatomega$.
    $\mathrm{Ad}$-equivariance by $N$ then uniquely determines
    $\bhatomega$ on the whole of $Q$.  A computation which is quite
    direct, only somewhat tedious to write down, then shows that the
    $\bhatomega$ thus defined really is a connection.

    The expression for the pullback of the curvature form follows
    directly by pulling back the structure equation $\boldsymbol{\hat
      R} = \D\bhatomega + \frac{1}{2} [\bhatomega \wedge \bhatomega]$.
  \end{proof}
\end{theorem}

As noted in the main text, this construction generalises the classical
situation of connections on the affine frame bundle of a manifold
(see, e.g., \cite [section III.3]{Kobayashi.Nomizu:1963}).  Note also
that in the case $\dim M = \dim N$, our construction may be seen as a
special case of Cartan geometry \cite{Sharpe:1997}---for the
application in the main text however, namely the consideration of
$\Barg$ bundles over Galilei manifolds, this is not satisfied.

\end{document}